\documentclass[prd,twocolumn,nofootinbib,showpacs]{revtex4-1}

\usepackage{amsfonts}
\usepackage{amsmath}
\usepackage{amssymb}
\usepackage{bm}
\usepackage{dcolumn}
\usepackage{graphicx}
\usepackage{graphics}
\usepackage[latin1]{inputenc}
\usepackage{latexsym}
\usepackage{rotating}
\usepackage{hyperref}
\usepackage[all]{hypcap} 
\usepackage{xspace} 
\usepackage[usenames,dvipsnames]{color}
\usepackage{mathrsfs}
\usepackage{subfigure}
\usepackage{aas_macros}
\usepackage{amsfonts}
\usepackage{booktabs}
\usepackage{siunitx}
\usepackage{appendix}

\usepackage{ulem}
\usepackage{outlines}
\usepackage{enumitem}
\setenumerate[1]{label=\Roman*.}
\setenumerate[2]{label=\Alph*.}
\setenumerate[3]{label=\roman*.}
\setenumerate[4]{label=\alph*.}

\definecolor {darkgreen}{rgb}{0.2,0.7,0.2}

\newcommand\be{\begin{equation}}
\newcommand\ba{\begin{eqnarray}}
\newcommand\ee{\end{equation}}
\newcommand\ea{\end{eqnarray}}

\newcommand\bw{\begin{widetext}}
\newcommand\ew{\end{widetext}}

\begin{document}
\title{Data Analysis Implications of Moderately Eccentric Gravitational Waves}

\author{Blake Moore}
\affiliation{eXtreme Gravity Institute, Department of Physics, Montana State University, Bozeman, MT 59717, USA.}

\author{Nicol\'as Yunes}
\affiliation{eXtreme Gravity Institute, Department of Physics, Montana State University, Bozeman, MT 59717, USA.}

\date{\today}

\begin{abstract} 
While the expectation is that the majority of gravitational wave events observable by ground-based detectors will be emitted by compact binaries in quasi-circular orbits, the growing number of detections suggests the possibility of detecting waves from binaries with non-negligible orbital eccentricity in the near future. 
Several gravitational wave models incorporate the effects of small orbital eccentricities ($e \lesssim 0.2$), but these models may not be sufficient to analyze waves from systems with moderate eccentricity. 
We recently developed a gravitational wave model that faithfully accounts for eccentric corrections in the moderate eccentricity regime ($e \lesssim 0.8$ for certain source masses) at 3rd post-Newtonian order.
Here we consider the data analysis implications of this particular waveform model by producing and analyzing posteriors via Markov Chain Monte Carlo methods. 
We find that the accuracy to which eccentricity and source masses can be measured can increase by 2 orders of magnitude with increasing eccentricity of the signal. 
We also find that signals with low eccentricity can be confidently identified as eccentric as soon as their eccentricity exceeds 0.008 (0.05) for low (high) mass systems, suggesting eccentric detections are likely to come first from low-mass systems.
We complete our analysis by investigating the systematic (mismodeling) error inherent in our post-Newtonian model, finding that for signals with a signal-to-noise ratio of 15, the systematic error is below the statistical error for eccentricities as high as 0.8 (0.5) for low (high) mass systems. 
We also investigate the systematic error that arises from using a model that neglects eccentricity when the signal is truly eccentric, finding that the systematic error exceeds the statistical error in mass for eccentricities as small as 0.02.
As a byproduct of this work we also present some new measures of the accuracy of our model, and investigate the efficiency of the model, showing that it can retain accuracy for large eccentricities while remaining efficient enough to be realistically useful for data analysis purposes.
\end{abstract}



\maketitle

\section{Introduction}
\label{intro}
We have now entered an era where gravitational wave (GW) detection is routine, enough so to compose a catalog of the 11 confirmed events \cite{2018arXiv181112907T} and explore the properties of populations of compact binaries \cite{2018arXiv181112940T}, such as mass and spin distributions. These GW detections provide information about the compact objects that emit them, but also they can be used to shed light on their formation scenarios, which helps further our knowledge of astrophysics and the universe at large. Currently only the masses, spins, distance, and sky orientation are modeled and measured, but there has been much work in producing models that incorporate orbital eccentricity \cite{2014PhRvD..90h4016H, Moore:2016qxz, 2019PhRvD..99l4008T, 2019arXiv190902143L, 2019CQGra..36r5003M, 2018arXiv180108542K}.

The traditional expectation is that gravitational wave observations with ground-based detectors should correspond to binaries in quasicircular orbits, because GW emission causes an eccentric binary to circularize. Recent studies \cite{2019arXiv190905466R, 2019arXiv190709384L} have searched for eccentric signals in the LIGO data and concluded that all of the current detections are consistent with binaries in quasicircular orbits. However, this is not completely inconsistent with the expectations of astrophysical scenarios, which suggest that there will be some small number of eccentric signals seen by ground-based detectors. With the GW catalog expected to expand considerably in the coming months and years, it is possible we will see at least one such event.

For binaries with detectable eccentricities emitting in the sensitivity range of ground-based detector networks, the main formation channels are dense stellar regions, such as globular clusters (GC) and the environment near supermassive black holes (SMBHs). Several recent works have incorporated post-Newtonian (PN) effects in population synthesis studies, focusing on binary formation in GCs and finding an increased number of binaries with eccentricity, with rates between 0.5 and 0.7 eccentric events per Gpc per year, and about 5$\%$ of all GWs in GCs having $e > 0.1$ when detectable by ground-based networks \cite{2018PhRvL.120o1101R, 2018PhRvD..97j3014S, 2014ApJ...784...71S, 2019ApJ...871...91Z}. About 0.5$\%$ of events near SMBH binaries with detectable mergers are expected to have $e > 0.5$ \cite{2012ApJ...757...27A, 2009MNRAS.395.2127O, 2014ApJ...781...45A}. For the space based Laser Interferometer Space Antenna (LISA), the expected number of eccentric sources considerably increases as they have not yet circularized due to GW emission \cite{2019ApJ...875...75F}.

With these distinct formation channels that can potentially assemble compact binaries in eccentric orbits comes the ability to constrain the formation mechanisms given a number of observations of eccentric events. In \cite{2019MNRAS.486..570T}, the authors claim that given tens of observations by LIGO one could discriminate between formation by the Kozai-Lidov mechanism and gravitational capture in GC. For LISA, only a handful of observations are needed to distinguish a population of events formed near SMBHs, and tens of events to distinguish between a population of binaries formed in the field versus in GCs \cite{2016PhRvD..94f4020N, 2017MNRAS.465.4375N}. Given the rich science potential of eccentric signals and the increasing likelihood of detecting such events, it is important to begin to investigate the data analysis implications and feasibility of models that incorporate eccentricity. 

Several studies have considered some of the data analysis implications of eccentric gravitational waves. In \cite{2013arXiv1310.8288F}, the authors study the systematic parameter error due to using quasi-circular templates to recover parameters from an eccentric signal using a Fisher analysis. Also recently in \cite{2019arXiv190911011R}, the authors produce hybrid eccentric PN-NR waveforms and again study the systematic error in parameter extraction from using circular templates on eccentric signals with a Bayesian methodology. The former found significant parameter error for low mass systems for signals with eccentricities near $10^{-2}$, and the latter found significant parameter error for higher mass systems for signals with eccentricities larger than $10^{-1}$. In \cite{2019ApJ...871..178G, Gond_n_2018}, the authors characterize some of the statistical errors associated with the various parameters of moderate and highly eccentric signals using a 1PN valid model and a Fisher analysis.

Previous studies were limited either by the computational expense or the accuracy of the eccentric model used. Here we carry out a fully Bayesian analysis of eccentric signals using Markov Chain Monte Carlo to produce the posterior distributions for the various parameters of eccentric systems. The key to this analysis is the use of the TaylorF2e model derived in \cite{2019CQGra..36r5003M}, which has been slightly modified to increase efficiency while retaining model accuracy. The model is inspiral only, valid to 3PN in the phasing, while keeping only the leading-order amplitude corrections. As in~\cite{2019CQGra..36r5003M}, we also illustrate that this model is accurate to eccentricities of $\sim 0.8$ for certain masses, and we also demonstrate that it is highly computationally efficient ($\sim$90ms per waveform evaluation). 

With this model in hand, we create simulated data using TaylorF2e and study the ability of a LIGO detector to recover parameters in both the moderate and low eccentricity regime. Generally, as the eccentricity of the signal is increased from 0.1 to $> 0.4$, the accuracy to which we can measure the eccentricity in the model improves by a factor of $10^{2}$, leading to statistical uncertainties on this parameter of order $10^{-5}$ ($10^{-3}$) for low (high) mass systems. We also find that the statistical uncertainty on the extraction of the  chirp mass decreases by a factor of 10 as the eccentricity of the signal increases. In the low eccentricity regime ($e < 0.1$), the eccentricity of the model becomes detectable when the eccentricity of the signal reaches values of about $\gtrsim 0.008$ ($\gtrsim 0.05$) for low (high) mass systems. This suggests that we might expect a larger number of statistically significant measurements of eccentricity from low mass compact binaries, unless these do not exist in Nature.

In addition to characterizing the measurement error sourced by detector noise, we also study the systematic error due to mismodeling in two different cases. We first study the recovery of parameters when fitting a circular waveform model to an eccentric signal. We find that there is significant error in the extraction of parameters when the eccentricities of the signal is as small as 0.01 (0.08) for low (high) mass systems. This underscores the importance of using eccentric models even when the eccentricity of the signal is expected to be very small. We then also studied the parameter error induced by inaccuracies in our eccentric model, such as that due to the truncation of expansions at a finite order. This error is studied by estimating the parameters in the frequency-domain, TaylorF2e model when the signal is a fully numerically-solved, 3PN, time-domain waveform. We here find that our model incurs parameter error that is smaller that the statistical error for eccentricities of at least $\lesssim 0.8$ in the low mass case, and $\lesssim 0.5$ in the high mass cases. This again points to the faithfulness of the model for parameter estimation in the moderate eccentricity regime.

The organization of the paper is as follows: Section \ref{sec:model} presents the model we use for our analysis, presenting its validity and efficiency. Section \ref{sec:bayes} reviews the Bayesian techniques we use in this work, and Section \ref{sec:large_ecc} presents and analyses parameter estimation in the high eccentricity regime, while Section \ref{sec:small_ecc} treats the low eccentricity regime. Sections \ref{sec:err_circ} and \ref{sec:err_model} study systematic parameter error from neglecting eccentricity and inaccuracy in our eccentric model, respectively. Section \ref{sec:conc} presents conclusions and points to areas to improve this study and possible interesting extensions. Throughout this work, we use geometric units (c = G = 1).

\section{Eccentric Model}
\label{sec:model}
In this section we summarize the 3PN-accurate, Fourier domain model that incorporates effects of periastron precession developed in~\cite{2018CQGra..35w5006M,2019CQGra..36r5003M}. This is the model that we will use to carry out our Bayesian analysis of binaries with moderate eccentricities. We implement minor modifications to this model that increases its computational efficiency, and so we also re-validate it against a fully numerical time-domain PN model. Lastly, we discuss the computational efficiency of generating the model, which has been recently added to the open-source \texttt{PyCBC} GW analysis package as \texttt{TaylorF2e}.

The model that we use for our Bayesian analysis is an optimized version of the TaylorF2e model derived and presented in \cite{2019CQGra..36r5003M}. In that paper we wrote our semi-analytic frequency-domain model schematically as
\begin{align}
\label{eq:spa}
\tilde{h}(f) &= A\left[ F_{+}S^2\sum_{s=1}^{\infty} G_s\sqrt{\frac{2\pi}{|s\ddot{l}|}}e^{i\psi_s} 
\nonumber \right.  \\ 
&+ Q\sum_{j=-1}^{\infty}N_j\sqrt{\frac{2\pi}{|j+2|\ddot{l}}}e^{i\psi^{-}_j}
\nonumber  \\ 
& \left. + Q^{\ast}\sum_{j=-\infty}^{-4}N_j\sqrt{\frac{2\pi}{|j+2|\ddot{l}}}e^{i\psi^{+}_j} \right] \,
\end{align}
where $Q$ is a function of the binary orientation relative to the detector frame, $A$ is an overall amplitude depending on the masses and distance to source, the $G_s$ and $N_j$ are amplitudes which control the strength of the different harmonics appearing in the sums that scale as $e^{|s|}$ and $e^{|j|}$ respectively (where here $e = e_{t}$ is the temporal PN eccentricity). In investigating techniques to optimize the model we found that we could neglect a significant number of harmonics appearing in the infinite sums of Eq.~\eqref{eq:spa} without losing much accuracy in the regions where the model was shown to be faithful to a fully numerical TaylorT4 model. 

Let us now make the parameter dependence and our optimizations more explicit. We neglect several harmonics in Eq.\eqref{eq:spa}, and rewrite the model in the form of
\begin{align}
\label{eq:spa_simp}
\tilde{h}(f) &= 10\pi \mathcal{A}\frac{\mathcal{M}_c}{\eta^{1/5}}\sum_{j=-1}^{15}N_j\sqrt{\frac{y^{-7}}{(j+2)(96 + 292e^2 + 37e^4)}}e^{i\psi^{-}_j} \,,
\end{align}
where $\mathcal{A}$ is an overall amplitude that depends on the sky angles, polarization angle and distance to source, $\eta = m_1 m_2/M^2$ and $\mathcal{M}_c = M \eta^{3/5}$ are the symmetric mass ratio and chirp mass respectively, with $M$ the total mass, and $m_1$ and $m_2$ the component masses, and $y = {\cal{O}}[(M n)^{1/3}]$ is our PN expansion parameter, with $n = 2 \pi/P$ the mean motion and $P$ the orbital period. The phases, $\psi^{-}_j$, are analytic functions of eccentricity given by
\begin{equation}
\label{eq:phases}
\psi^{-}_j = 2\pi f \left[t(e) - t_c\right] - j \left[l(e) - l_c\right] - 2 \left[\lambda(e) - \lambda_c\right] - \frac{\pi}{4}\,, 
\end{equation} 
where $\lambda$ and $l$ are angles associated with the orbit and $t - t_c$ is the time to coalescence. These phase functions depend on the masses and mass ratio, and are series expansions in $e$. In this work, we use both the version of the waveform that keeps more terms in eccentricity in these phase functions (TaylorF2e+), and the version which keeps less terms (TaylorF2e-). For a summary of how many terms each keeps, see Table I of \cite{2019CQGra..36r5003M}.

In order to compute the model at a given frequency, one must numerically invert the stationary phase condition given a Fourier frequency for the corresponding eccentricity, which is then used in the model. The stationary phase condition reads
\begin{equation}
\label{eq:stat_ph_cond}
2\pi f = j\dot{l}(e)+2\dot{\lambda}(e) \, ,
\end{equation}
where an overdot denotes the time derivative. For much more detail on the model derivation, see \cite{2019CQGra..36r5003M}. 

A further optimization that we implement is an adaptive scheme to sample the different $j$ harmonics appearing in the sum in Eq.~\eqref{eq:spa_simp} in frequency. Any given harmonic $j$ begins emitting at a lower frequency $f_{0,j}$ obtained through Eq.\eqref{eq:stat_ph_cond} as $2\pi f_{0,j} = j\dot{l}(e_0) + 2\dot{\lambda}(e_0)$. Here $e_0$ is the eccentricity at an initial time and separation, where recall that we cannot assign $e_0$ to a single initial GW frequency because the signal is composed of many harmonics. Any given harmonic $j$ ends emission at $2\pi f_{\rm fin, j} = j\dot{l}(e_{\rm fin}) + 2\dot{\lambda}(e_{\rm fin})$, where $e_{\rm fin}$ is chosen to correspond to some truncation condition. We truncate our waveforms when the periastron velocity reaches one-third the speed of light. Thus, higher harmonics emit in a wide frequency band, but inspection of the $N_j$ reveals that they rapidly tend to zero as $e$ decreases (see Fig.~6 of \cite{2019CQGra..36r5003M}). As a result, it is inefficient to waste computational resources computing the higher harmonics once their contribution to the signal becomes negligible. We choose to stop sampling the $j^{\rm th}$ harmonic when $e$ satisfies $e \leq (j-1)/28$ or $e < e_{\rm fin}$, whichever is satisfied first (a choice that can be made by inspection of Fig.~6 of \cite{2019CQGra..36r5003M}). We translate this condition on the eccentricity into a starting and ending frequency for each harmonic using Eq.\eqref{eq:stat_ph_cond}.

\begin{figure*}[htp]
\includegraphics[clip=true,angle=0,width=0.475\textwidth]{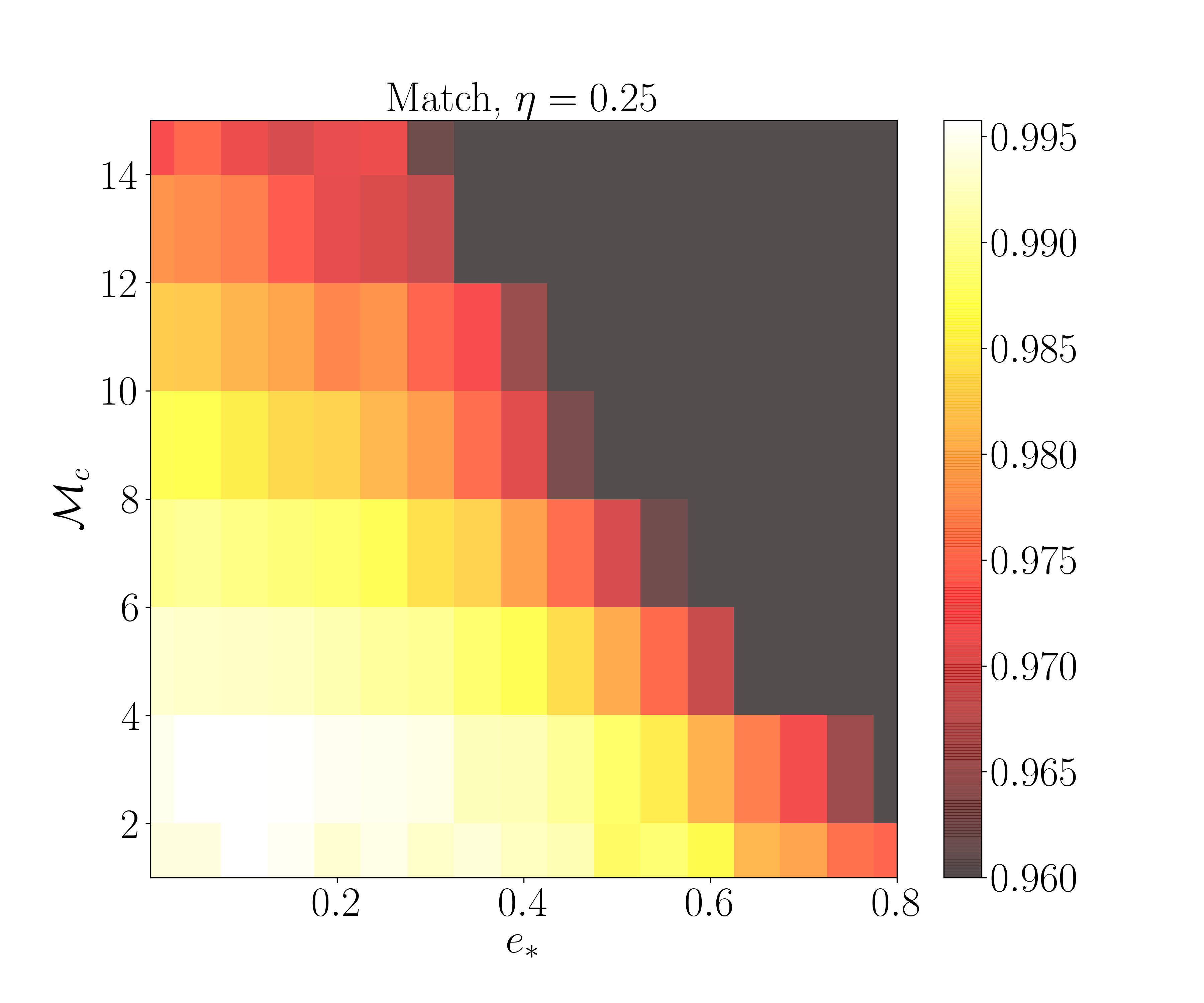}
\includegraphics[clip=true,angle=0,width=0.475\textwidth]{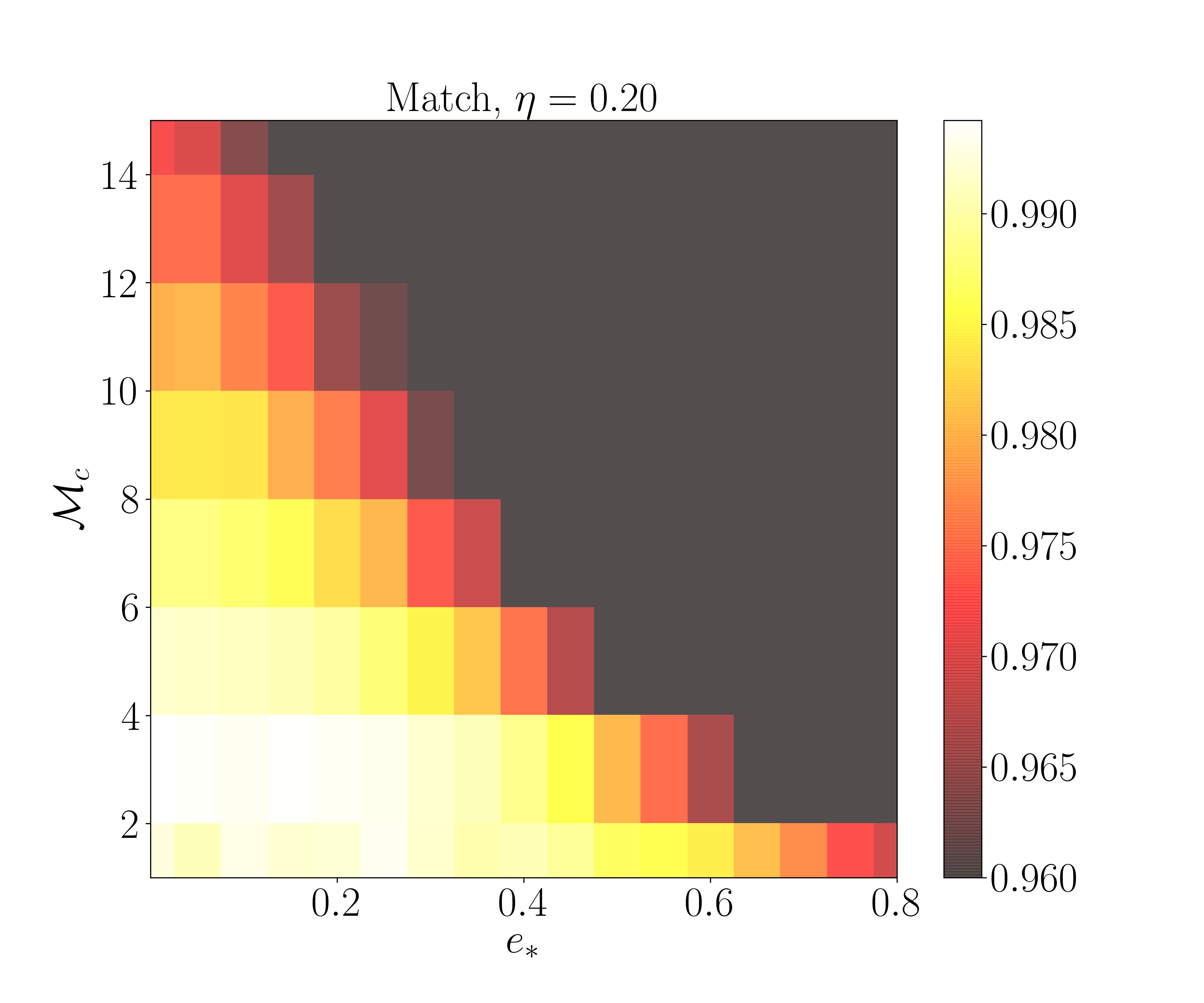}
\caption{\label{fig:match_f2e+} The match between the TaylorF2e+ model and a fully numerical, time-domain model as a function of the chirp mass and $e_{\ast}$ for $\eta = 0.25$ (left) and $\eta = 0.2$ (right). Values below $96\%$ are excluded. As the mass becomes unequal, the match deteriorates. The model is more faithful for lower mass systems.}
\end{figure*}
We must be careful in the way that we parameterize the orbital eccentricity in our model. In \cite{2019CQGra..36r5003M}, we parameterized the model in terms of $e_0$ and $p_0$, which were to be understood as the eccentricity and semi-latus rectum at some initial time. This choice was made because we were taking the Fourier transform of a time-domain signal, and subsequently using it to validate an  approximate frequency-domain model. The time-domain signal had to be started at a given initial time $t_{0}$ with some initial eccentricity $e_{0}$ and semi-latus rectum $p_{0}$. Each harmonic of the GW signal would then emit in some frequency range that depended on the harmonic index $j$, as described above. However, these choices came with the physical interpretation that the binary had to spontaneously begin emitting at $t_0$, and there was no emission before this time. In reality, binary systems are likely to have been around long before the fiducial time $t_{0}$ that we chose, and thus, all harmonics should have been emitting significantly from initial frequencies much lower than the lower frequency cutoff of ground-based detectors. Thus, this time around, we would like to construct the model in such a way that we choose some reference eccentricity and reference separation, and then compute all the relevant GW emission self-consistently, some of which will likely come from earlier times when the orbit had a higher eccentricity and a larger separation. 

What is the best way to choose this reference eccentricity and separation? What is commonly done is to choose the ``eccentricity at a GW frequency of 10Hz". However, this description is problematic because, at any given time and eccentricity, an eccentric binary emits GWs at multiple frequencies. What is typically meant by the previous statement is that one chooses the ``eccentricity at an orbital frequency of 5 Hz," since the only harmonic that is non-vanishing in the circular limit emits at twice the orbital frequency. This amounts to fixing the eccentricity at some reference semi-major axis by Kepler's third law. While this is a sensible choice for small eccentricities, for large eccentricities this choice becomes troublesome. While the semi-major axis may be large, for large eccentricities the corresponding pericenter distance can be very, very small, and sometimes so small than the PN approximation should not be trusted. For example, if we fix the reference semi-major axis to be $50M$ (corresponding to a reference orbital frequency of about 5 Hz for a $(10,10)M_{\odot}$ system) and increase the orbital eccentricity from 0.1, to 0.3, 0.6, and 0.9, the reference pericenter distance becomes 45$M$, 35$M$, 20$M$, and 5$M$. 

This way of choosing a reference eccentricity is clearly inappropriate, since it can correspond to an orbit with such small pericenter distances that the PN approximation breaks down. Instead, we here choose to parameterize our model by \emph{specifying the eccentricity at the semi-latus rectum that corresponds to the separation of a circular binary emitting GWs at 10 Hz}. With this choice, if we now fix the semi-latus rectum to be 50$M$ and increase the orbital eccentricity from 0.1, to 0.3, 0.6, and 0.9, then the corresponding pericenter distances are 45$M$, 38$M$, 31$M$, and 26$M$. This choice has the advantage that we are not specifying orbits outside the PN regime of validity, and as the eccentricity is decreased, this choice will nearly match the conventions of early models. We will refer to this reference eccentricity parameter as $e_{\ast}$.

Given this new parameterization, we must now re-validate the model against a fully numerical, time-domain signal. In order to achieve this, we generate a time-domain model whose initial conditions ($e_0, p_0$) are determined by the separation and eccentricity sampled by our TaylorF2e's highest harmonic at 10 Hz. This time-domain model and our techniques for calculating the match are described in detail in \cite{2019CQGra..36r5003M}. Figure \ref{fig:match_f2e+} shows the match between our frequency-domain model TaylorF2e+ and the fully numerical, time-domain model as a function of the chirp mass and $e_{\ast}$. The match deteriorates as the mass and eccentricity is increased, and as $\eta$ is decreased. In Appendix \ref{app:validation}, we present plots of the fitting factors as well as match for the model that keeps less terms in the eccentric expansions in the phase TaylorF2e-. There is a significant difference in the regime of validity between the two frequency-domain models when the mass is low and the eccentricity is high. 

We conclude this section with a note about the computational performance of the model. In order to gauge the speed of generating our model, we draw parameters from uniform random variables. We draw the chirp mass from $U[0.8, 15]$, $\eta$ from $U[0.15, 0.25]$, $e_{\ast}$ from $U[10^{-6}, 0.8]$, and $\mathcal{A}$ from $U[10^{-16}, 10^{-20}]$. We make $10^6$ draws and calculate the corresponding frequency response with a frequency resolution of $0.015625$Hz. The average time to compute the waveform in the parameter space described above is $61.4$ms per TaylorF2e- waveform evaluation and $94.8$ms per TaylorF2e+ waveform evaluation. This timing study was carried out on a 2.20GHz Intel Xeon E5-2630 v4 CPU.

\section{Bayesian Framework}
\label{sec:bayes}
In this section we provide a short summary of the technical details and techniques used to investigate the data analysis implications of moderately eccentric GWs. We use an MCMC algorithm to produce samples of the posterior given advanced LIGO at design sensitivity, assuming a single detector, and maximizing the likelihood over extrinsic parameters.
\begin{figure*}[htp]
\includegraphics[clip=true,angle=0,width=0.475\textwidth]{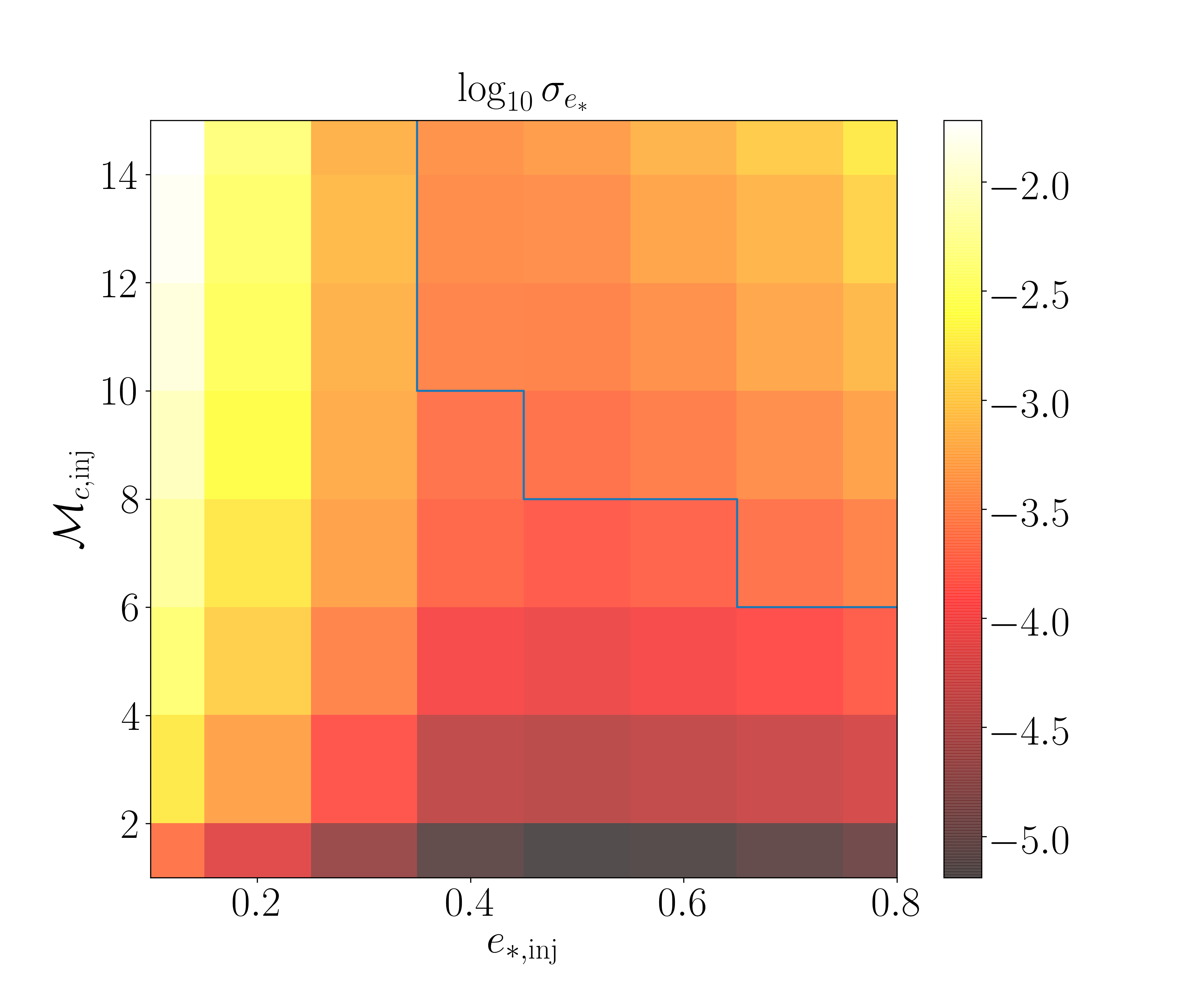}
\includegraphics[clip=true,angle=0,width=0.475\textwidth]{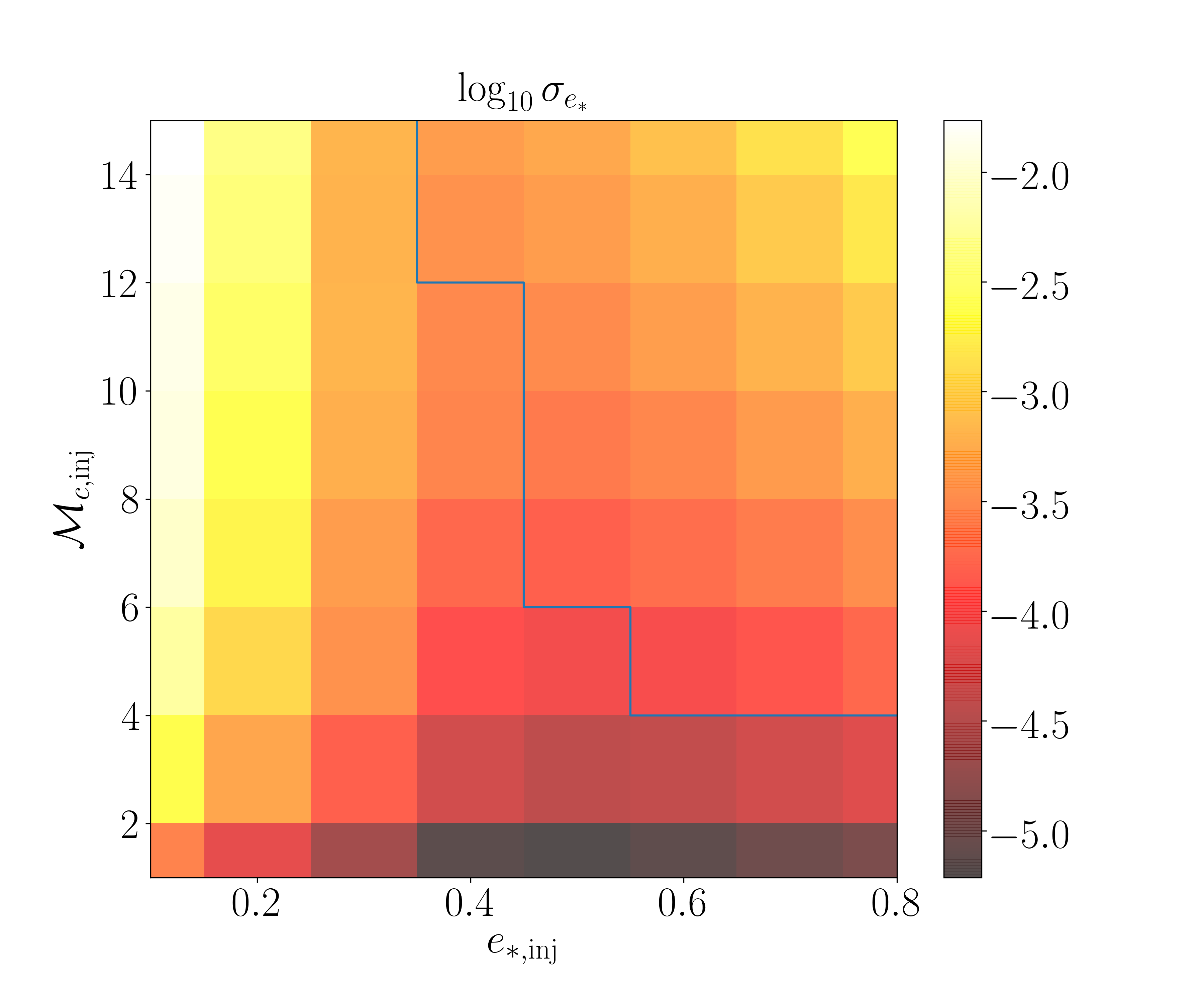}
\caption{\label{fig:sigma_heat} Log base 10 of the standard deviation of the marginalized posterior of eccentricity for different injected eccentricities $(e_{\ast, \rm inj})$ and chirp masses $(\mathcal{M}_{c, \rm inj})$, with $\eta_{\rm inj} = 0.25$ on the left panel and $0.20$ on the right panel, and with an SNR of 15. To the right of the blue line on the left panel the systematic error from mismodeling exceeds $\sigma_{e_{\ast}}$. Observe that the accuracy to which the eccentricity can be measured can exceed $10^{-5}$ for moderately eccentric and low mass systems.}
\end{figure*}

In order to answer questions about how well we can recover the parameters of the waveform model described in Section \ref{sec:model}, we employ Bayesian inference and obtain the posterior distribution $p(\bold{\theta}|\bold{d}, M)$. The posterior is the probability of the parameters ($\theta$) given the data ($\bold{d}$) and the model ($M$). This probability distribution is used to glean information about the recovered parameters, such as credible intervals, the mode, mean, and median. In order to work with and visualize the posterior in terms of a single parameter, we often work with the marginalized posterior, given by 
\begin{equation}
\label{eq:marg_post}
p(\theta^{j}) = \int p(\bold{\theta}|\bold{d}, M) \prod_{k \neq j} d\theta^{k} \, .
\end{equation}
In our case, we have reduced the model to four intrinsic parameters and three extrinsic parameters, $\bold{\theta} = \lbrace \ln M_c, \eta, e_{\ast}, \ln \mathcal{A} , \lambda_c, l_c, t_c \rbrace$, because we assume the binary is non-spinning, and because we consider a single detector with which it is not possible to obtain sky localization information.

In order to obtain the posterior distribution we have written our own Markov Chain Monte Carlo (MCMC) algorithm \cite{gilks1995markov}. In short the MCMC is a stochastic process whose random walk in parameter space produces the posterior distribution. In order to take steps, the MCMC requires the computation of the likelihood $\mathcal{L}(\bold{h})$, given by 
\begin{equation}
\label{eq:likelihood}
\mathcal{L}(\bold{h}) \sim \exp \left\lbrace - \frac{1}{2}(\bold{d} - \bold{h}|\bold{d} - \bold{h})\right \rbrace \, ,
\end{equation} 
where the inner product between a signal $h_1(t)$ and $h_2(t)$ is given by 
\begin{equation}
\label{eq:prod}
(h_1|h_2) = 4\text{Re}\int \frac{\tilde{h}_1^{*}(f) \tilde{h}_2(f)}{S_n(f)}df \, ,
\end{equation}
which also defines the signal to noise ratio (SNR) of a signal $h$:
\begin{equation}
\label{eq:SNR}
\rm SNR^2 = (\bold{h}|\bold{h}) \, .
\end{equation}
\begin{figure}[htp]
\includegraphics[clip=true,angle=0,width=0.475\textwidth]{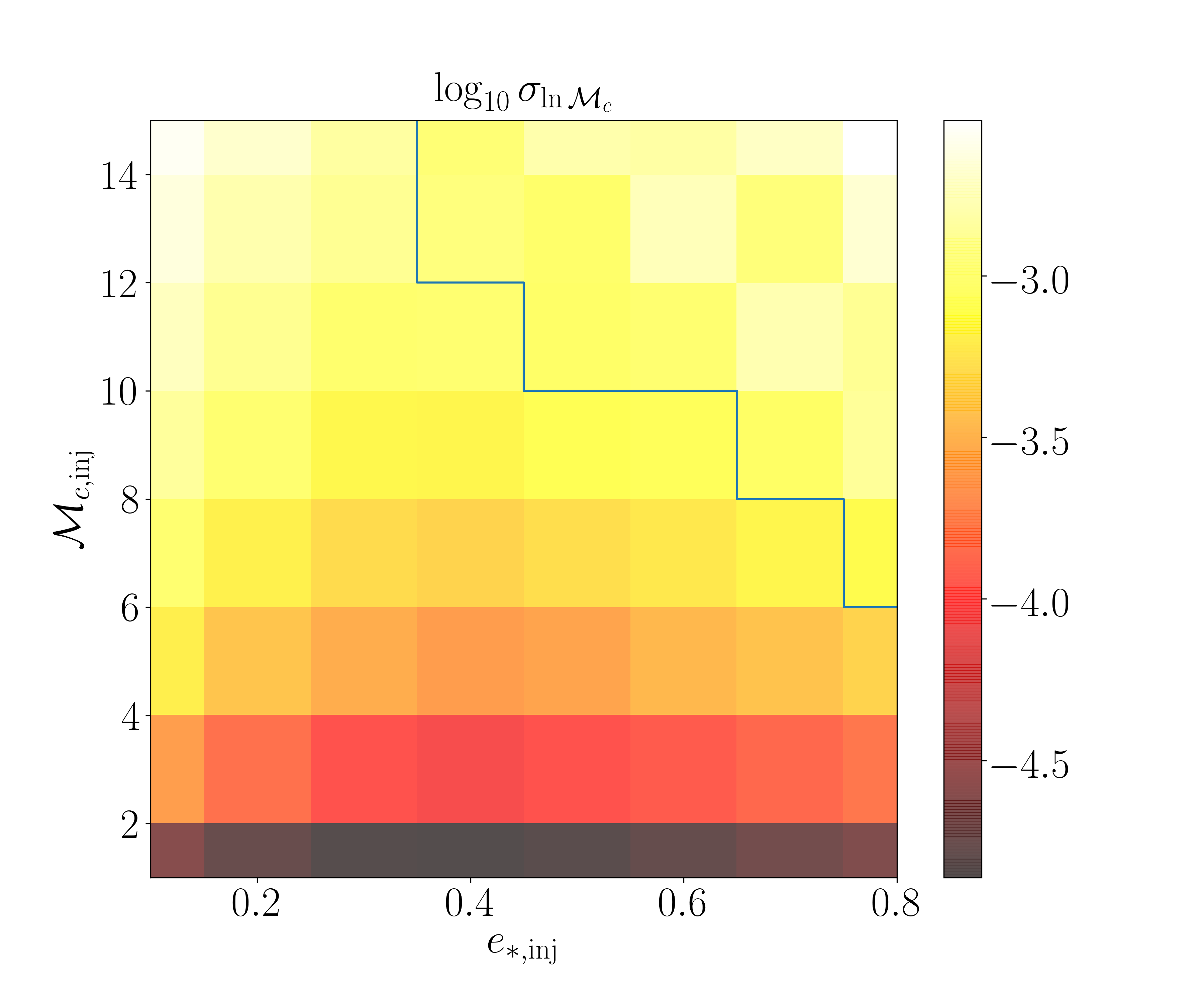}
\includegraphics[clip=true,angle=0,width=0.475\textwidth]{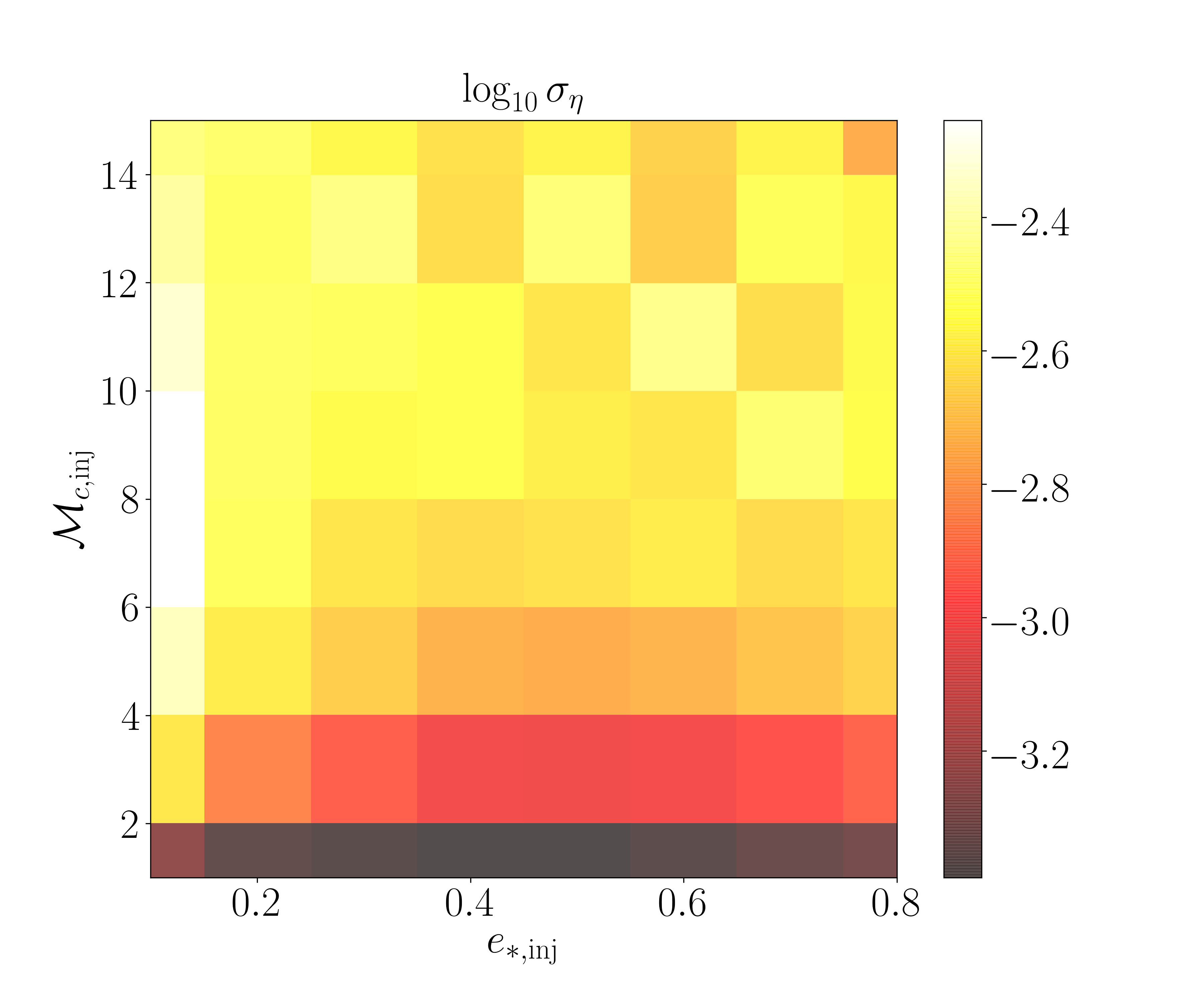}
\caption{\label{fig:sigma_mc} Same as Fig.~\ref{fig:sigma_heat} but for the $\ln \mathcal{M}_c$ (top) and $\eta$ (bottom) parameter. Observe that the accuracy to which both the chirp mass and reduced mass ratio can be measured improves for moderately eccentric signals.}
\end{figure}
In Eq. \eqref{eq:prod}, $\text{Re}$ stands for the real operator, $*$ denotes complex conjugation, the overhead tilde stands for the Fourier transform, and $S_n(f)$ is the spectral noise density of the detector. In this work we use spectral noise of aLIGO at designed sensitivity (zero-detuned, high-power), and assume stationary, Gaussian noise \cite{2017CQGra..34d4001A} when computing the likelihood. We consider one detector with no noise injection. In addition, we maximize the inner product appearing in the likelihood over the parameters $\lambda_c$, $l_c$, and $t_c$ which arise in the phases shown in Eq. \eqref{eq:phases} as harmonic dependent phase shifts and an overall time shift. This maximization procedure differs from the analogous procedure in the circular case, and was derived and explained in Sec.~III of \cite{2018CQGra..35w5006M}.

In order to complete the description of our MCMC algorithm, we must still specify the priors and the proposal distribution from which we draw in order for the algorithm to propose steps in parameter space. Our prior distributions are uniform flat on all parameters: $\ln \mathcal{M}_c \in [-0.223, 2.708]$, $\eta \in [0.15, 0.25]$, $e_{\ast} \in [10^{-6}, 0.8]$, and $\ln A \in [-36.8414, -46.0517]$. The choice of priors in chirp mass corresponds to the type of sources we explore here. Our restrictive choice of lower bound on the prior for $\eta$ is a consequence of eccentric models suffering from inaccuracy as $\eta$ becomes small (see Figure \ref{fig:match_f2e+} and Figures 7 and 10 of \cite{2019CQGra..36r5003M}). Likewise our choice of upper bound on $e_{\ast}$ represents the highest eccentricity we expect the model to be faithful for (for certain masses). Our proposal distribution consists of a mixture of Fisher matrix proposals, draws from the prior, and differential evolution. Additionally, we use parallel tempering to ensure that the entire parameter space is explored and efficiently sampled. We have checked and ensured that the hottest chains are exploring the entire parameter space. In order to determine convergence of the chains, we have verified that producing more samples of the posterior does not qualitatively affect our results.  

\section{How Well Can We Measure Large Eccentricities?}
\label{sec:large_ecc}
\begin{figure*}[htp]
\includegraphics[clip=true,angle=0,width=0.475\textwidth]{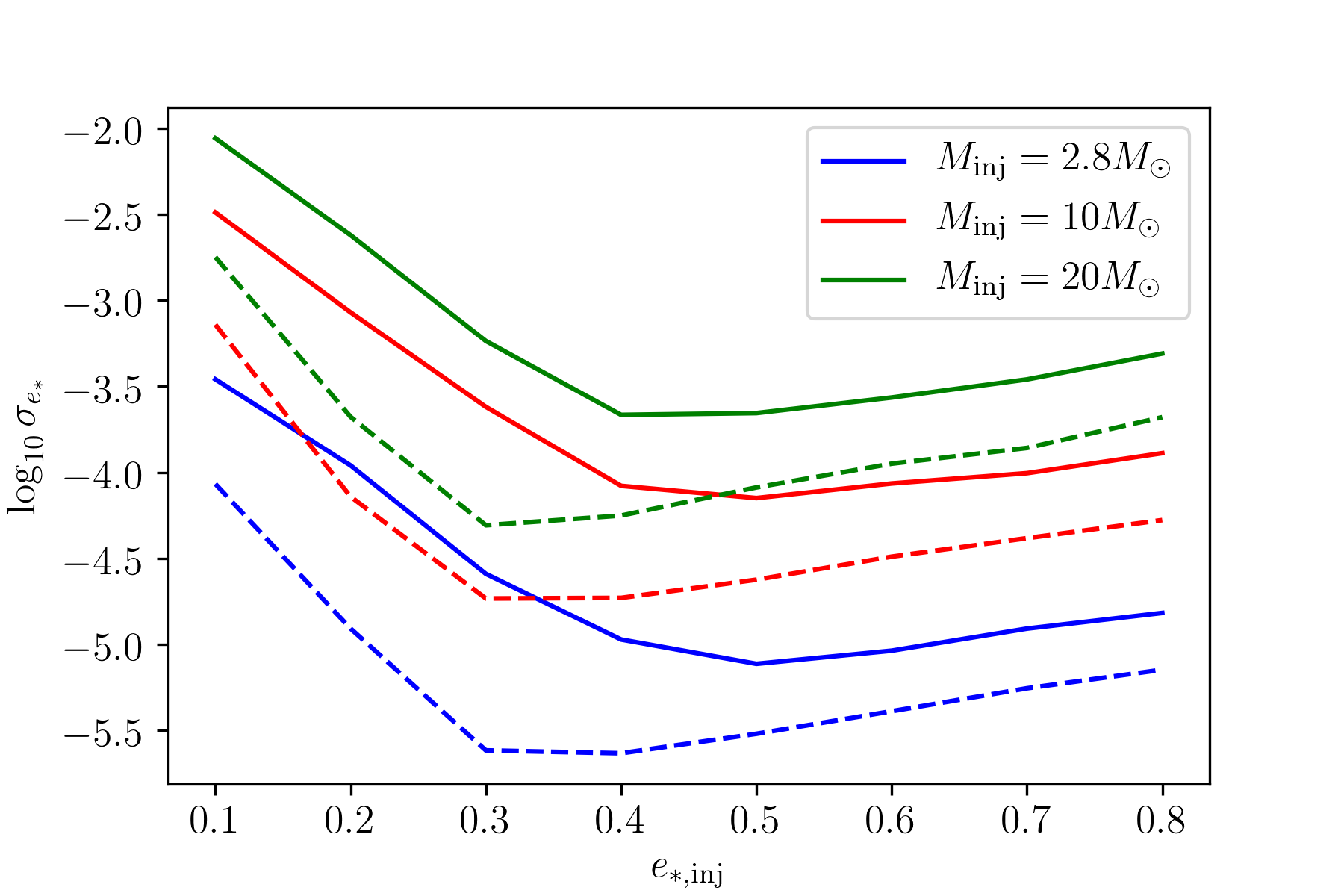}
\includegraphics[clip=true,angle=0,width=0.475\textwidth]{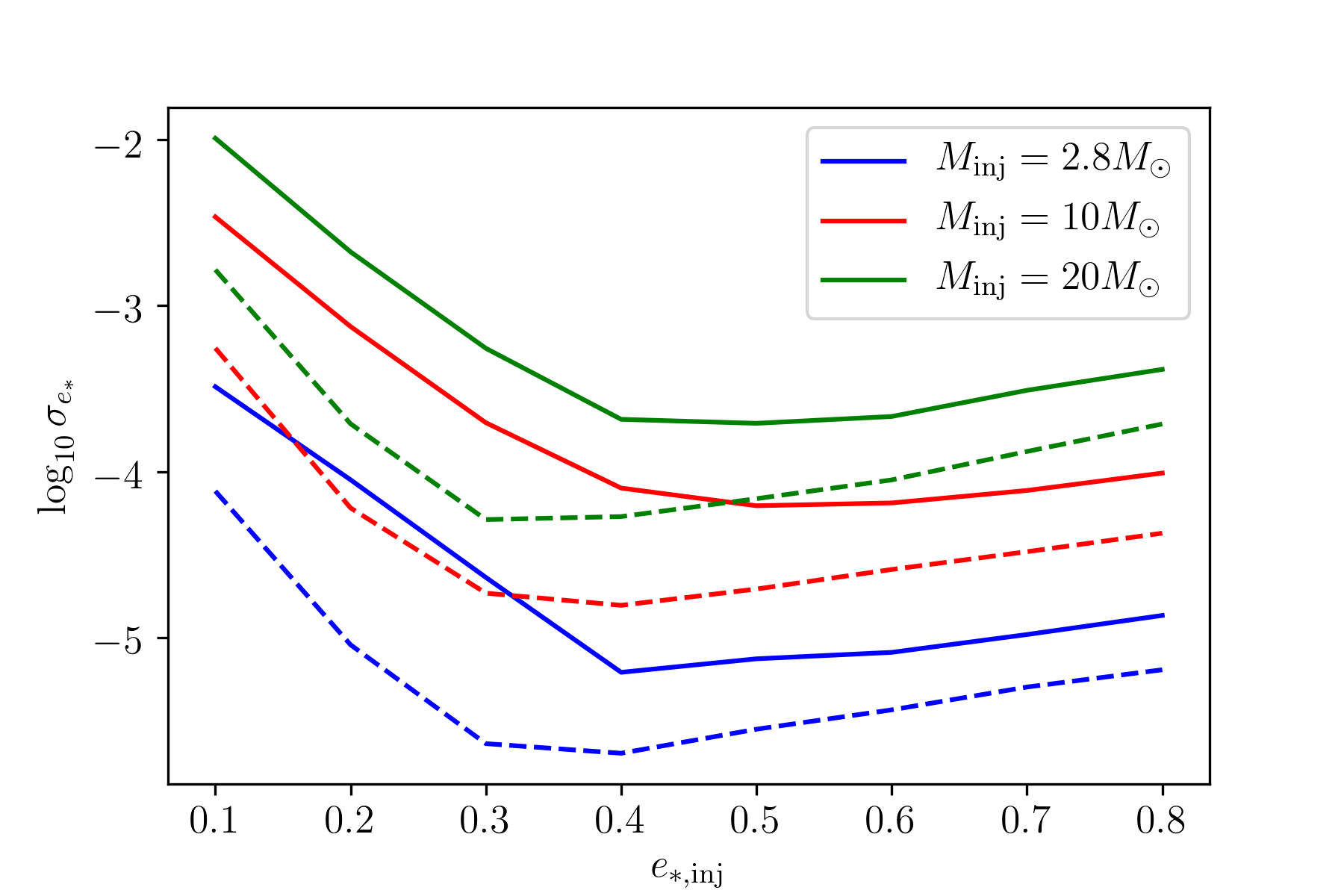}
\caption{\label{fig:var_snr} Log base 10 of the standard deviation of the marginalized posterior of eccentricity, $\sigma_{e_{\ast}}$ for different injected eccentricities holding the injected total mass ($M_{\rm inj}$) fixed. The signal SNR is 15 (solid) and 30 (dashed), while $\eta_{\rm inj}$ is 0.25 (left panel) and 0.20 (right panel). Observe that the accuracy to which the eccentricity can be measured improves with the injected eccentricity of the signal, but it reaches a saturation point at moderate injected eccentricities due to covariances (see Fig.~\ref{fig:cov}).}
\end{figure*}
\begin{figure}[htp]
\includegraphics[clip=true,angle=0,width=0.475\textwidth]{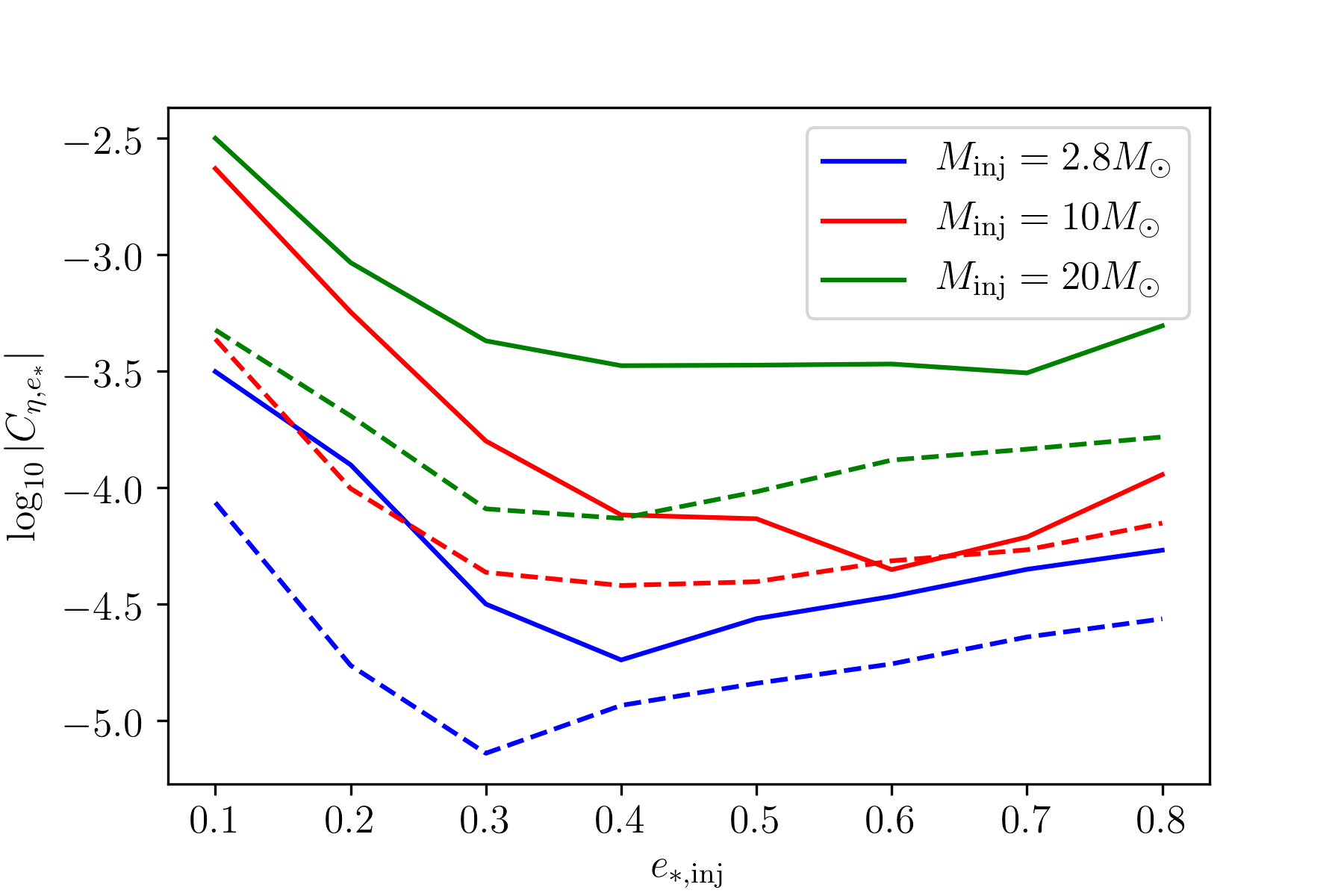}
\caption{\label{fig:cov} Log base 10 of the absolute value of the covariance between $\eta$ and $e_{\ast}$ for various total masses as a function of the injected eccentricity. The signal SNR is 15 (solid) and 30 (dashed). Observe that the covariance between these parameters improves with the injected eccentricity of the signal, until it saturates at moderate injected eccentricity, presenting a patter similar as that found for the variance of the eccentricity in Fig.~\ref{fig:var_snr}.}
\end{figure}
In this section we explore how well we can estimate the eccentricity of the model when the eccentricity of the injected signal is larger than 0.1. We consider two sources of parameter estimation error. There is a statistical error in parameter estimation due to covariances between model parameters and detector noise -- an error that will obviously depend on the SNR. There is also a systematic error in parameter estimation due to the template not being an exact representation of the true signal -- an error that will be independent of the SNR. We here seek to explore the measurement of eccentricity in the context of both errors. In producing statistical errors, we employ our TaylorF2e- model and have verified that the results are qualitatively unchanged when using the TaylorF2e+ model. However, to gauge systematic error we require a more faithful model and instead use the TaylorF2e+ model. 

Figure \ref{fig:sigma_heat} shows $\sigma_{e_{\ast}}$ (i.e.,~the standard deviation in the samples of the marginalized posterior on the parameter $e_{\ast}$) as a function of the injected chirp mass ($\mathcal{M}_{c, \rm inj}$) and eccentricity ($e_{\ast, \rm inj}$). We present results for two different mass ratios and all signals have an SNR of 15. In the top panel, we have added a line which indicates that, to the right (for larger injected eccentricities), the systematic error from mismodeling is greater than $\sigma_{e_{\ast}}$. Our technique for obtaining this systematic error is explained in Sec.~\ref{sec:err_model}, where we also provide analogous plots for this error. If we assume that the marginalized posterior in terms of eccentricity is Gaussian (an appropriate approximation according to the posteriors plotted in Fig.~\ref{fig:corner}), then the 1-sigma region indicates a region of 68$\%$ confidence. Thus, when the systematic error exceeds 1-sigma, our systematic error causes the measurement to be outside of this 68$\%$ confidence region. Although not shown in this figure, as the SNR is increased, the statistical error also decreases, and thus, we will eventually require more faithful modeling for higher mass systems with large eccentricities and higher SNRs. 

Let us now make note of some of the features of the statistical error shown in Fig.~\ref{fig:sigma_heat}. Overall, the eccentricity can be extracted to better than $10\%$, with typical values that go down all the way to $10^{-5}$ for low mass systems. The fact that as the injected chirp mass decreases, $\sigma_{e_{\ast}}$ becomes smaller is sensible. The signal contains more GW cycles for lower mass systems, and thus, the likelihood is more sensitive to changes in model parameters. While the difference in $\sigma_{e_{\ast}}$ is nearly indistinguishable by eye when $\eta_{\rm inj}$ is changed from $0.25$ to $0.20$ (from the left to the right panel), we have verified that $\sigma_{e_{\ast}}$ typically has a fractional difference of about 0.06 between the two cases.

Figure \ref{fig:sigma_mc} shows the standard deviation of the marginalized posterior of $\ln \mathcal{M}_c$. Again, the blue line indicates the threshold for when the systematic error exceeds 1-sigma. This line is slightly less restrictive than the systematic error in eccentricity. We see that, as the injected eccentricity is increased, the standard deviation decreases compared to the low eccentricity case. This suggests that moderately eccentric signals could provide better constraints on the chirp mass of the emitting system. The result for the unequal mass case is nearly indistinguishable by eye, so we do not include it here. We also include in Fig.~\ref{fig:sigma_mc} an analogous plot for the dependence of the statistical uncertainty in the symmetric mass ratio, $\eta$, as a function of injected chirp mass and eccentricity. For the low mass sources, this uncertainty decreases by about half an order of magnitude as the injected eccentricity is increased. For $\ln \mathcal{A}$, there is no trend in the statistical error with respect to eccentricity. 

Figure \ref{fig:var_snr} plots again $\log_{10}\sigma_{e_{\ast}}$, but this time holding the injected total mass fixed and changing the SNR from 15 (solid lines) to 30 (dashed lines). As expected, $\sigma_{e_{\ast}}$ decreases as the SNR is increased. However, we notice a counterintuitive result: $\sigma_{e_{\ast}}$ initially decreases as $e_{\ast, \rm inj}$ is increased, but hits a certain value of $e_{\ast, \rm inj}$, and then begins to slowly increase as $e_{\ast, \rm inj}$ continues to increase. One explanation for this behavior is that a covariance between $\eta$ and $e_{\ast}$ follows a similar trend. Figure \ref{fig:cov} shows $\log_{10}|C_{\eta, e_{\ast}}|$ for the same case as the top panel of \ref{fig:var_snr}. This covariance forces the statistical error on the marginalized posteriors of eccentricity to widen.
 
\section{How Well Can We Measure Small Eccentricities?}
\label{sec:small_ecc}

\begin{figure}[htp]
\includegraphics[clip=true,angle=0,width=0.475\textwidth]{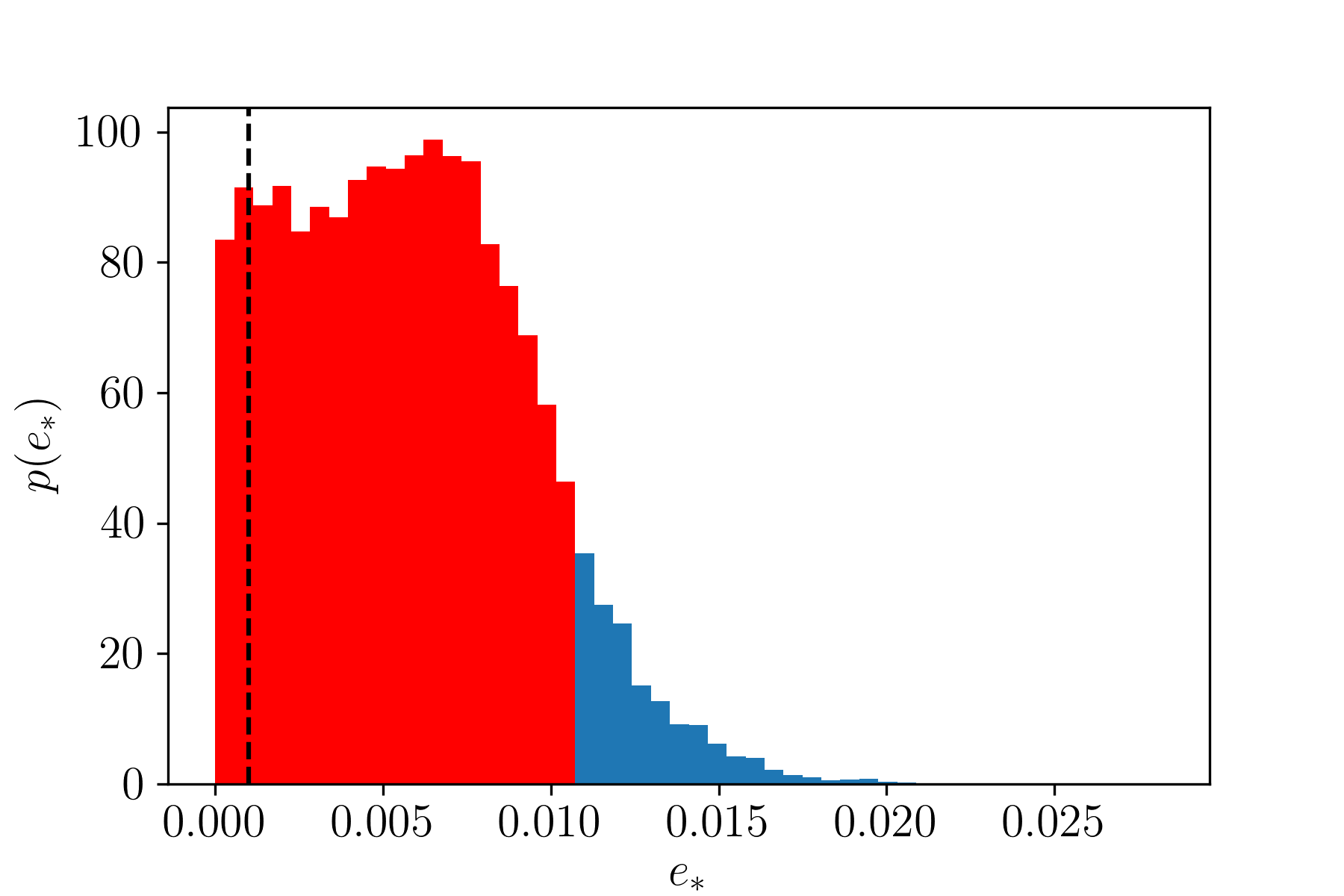} 
\includegraphics[clip=true,angle=0,width=0.475\textwidth]{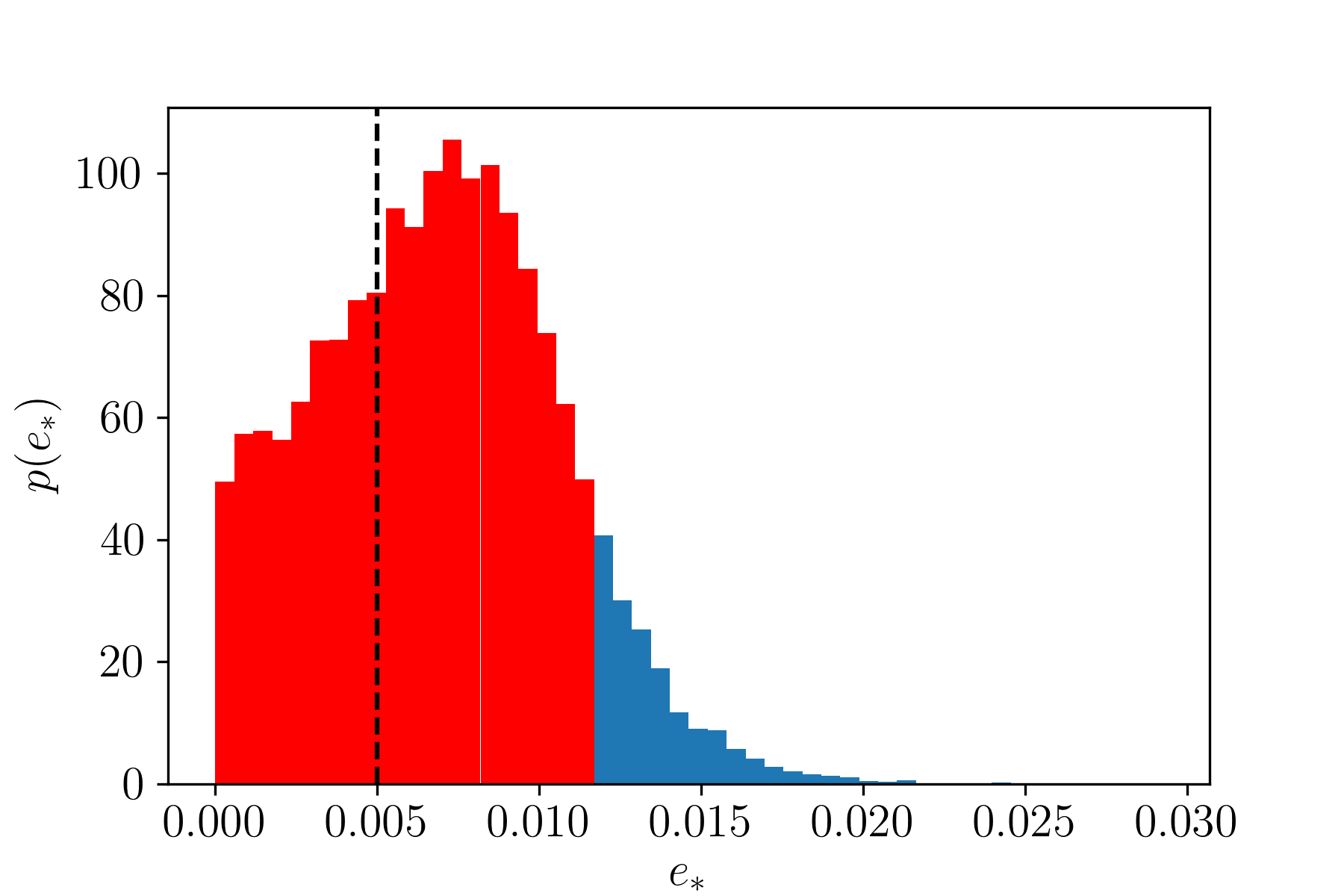} 
\includegraphics[clip=true,angle=0,width=0.475\textwidth]{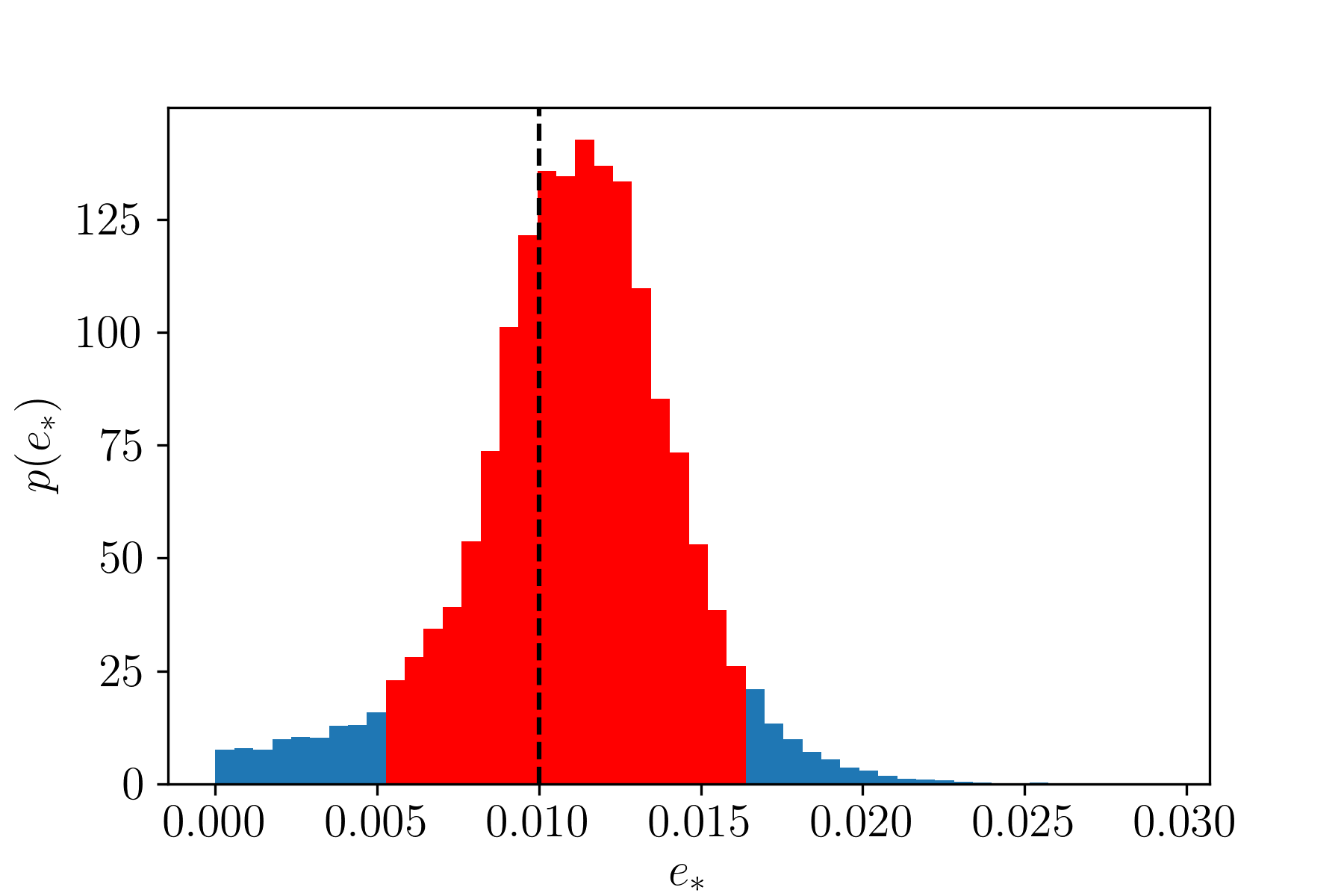}
\caption{\label{fig:low_e_posts} Marginalized posteriors on eccentricity for a $(1.4, 1.4)M_{\odot}$ mass injection with injected eccentricity of 0.001 (top), 0.005 (middle), and 0.01 (bottom). In red we have over-plotted the 90$\%$ confidence region, with the dashed line indicating the injected value of eccentricity. In the lowest eccentricity injection, we see that, in the region near zero eccentricity, the posterior is effectively sampling the flat prior. By 0.005 eccentricity, the posterior is beginning to have a more defined peak away from zero, and by 0.01 eccentricity, there is an even more well defined peak with zero eccentricity excluded from the 90$\%$ confidence region.}
\end{figure}
\begin{figure*}[htp]
\includegraphics[clip=true,angle=0,width=0.475\textwidth]{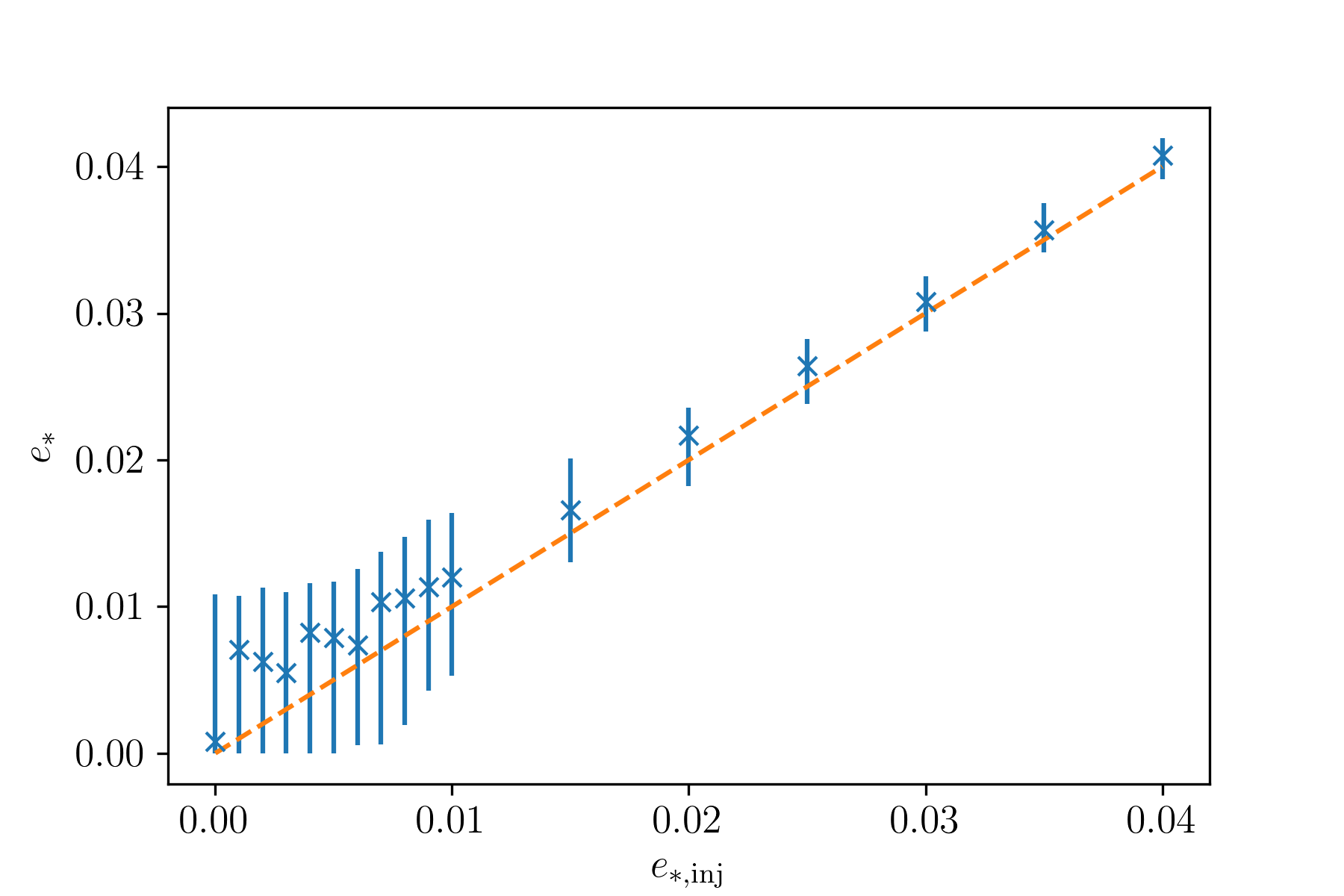}\includegraphics[clip=true,angle=0,width=0.475\textwidth]{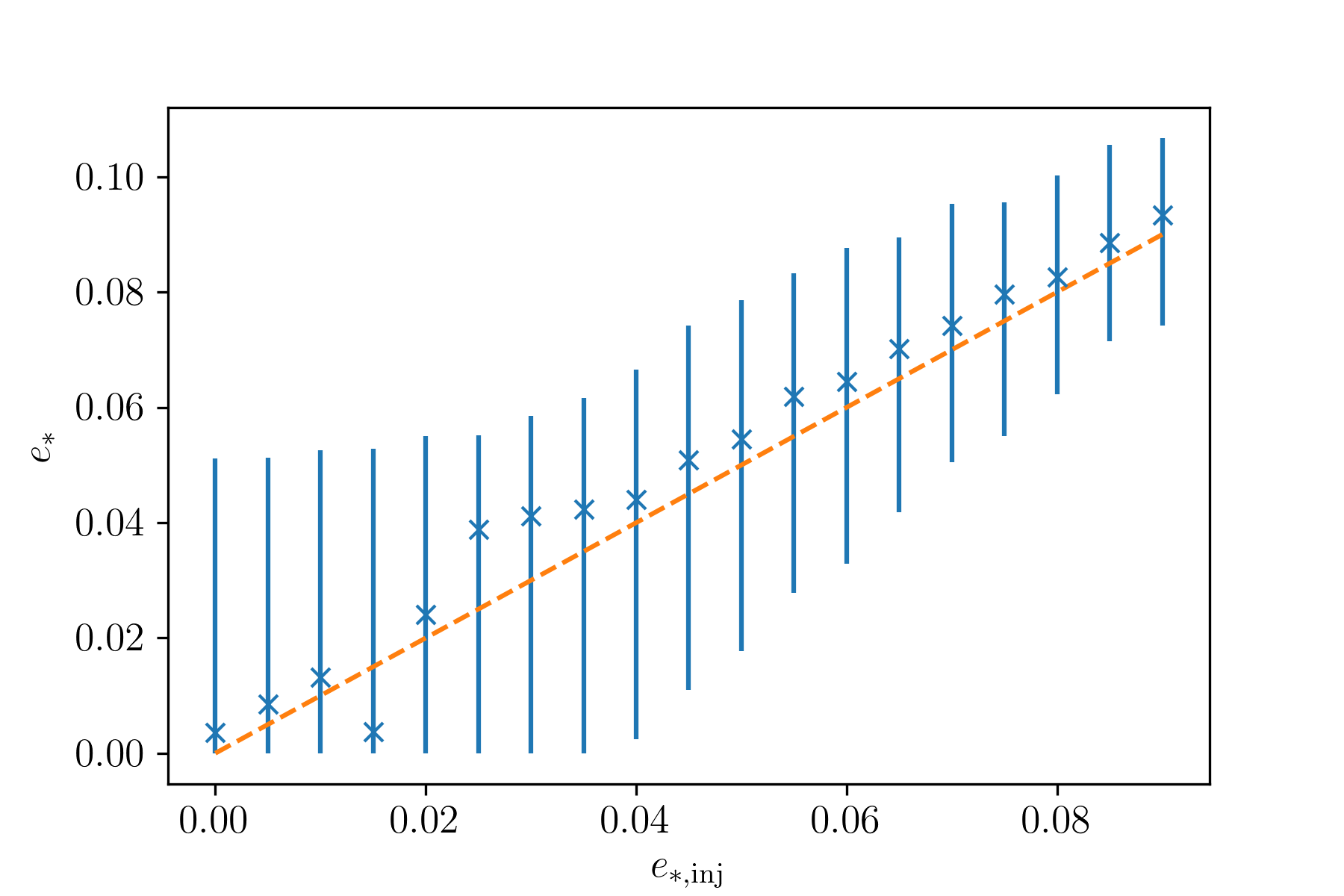}
\caption{\label{fig:low_e} A characterization of the marginalized posterior of eccentricity, $p(e_{\ast})$, when injected eccentricity is small. The vertical line indicates the 90$\%$ minimal area confidence interval, and the ``x" indicates the peak of the marginalized posterior. The dashed line indicates where $e_{\ast,\rm inj} = e_{\ast}$. We show results for a $(1.4, 1.4)M_{\odot}$ mass injection (left) and a $(10,10)M_{\odot}$ mass injection (right). For the low (high) mass injection, the 90$\%$ confidence interval begins to exclude 0 eccentricity around $e_{\ast} = 0.008 \, (0.05)$. The SNR of the injected signals are 15.}
\end{figure*}

In this section we study the ability to measure the eccentricity parameter of our model when the injected eccentricity of the signal is small. 
The most rigorous method to determine whether one can distinguish an eccentric model from a circular model is to compute the Bayes factor or Odds ratio. As the main focus of this work is to characterize measurement in the high eccentricity regime, we instead choose to investigate the small eccentricity regime more qualitatively, by characterizing the marginalized posteriors of eccentricity when the injected eccentricity is small. In particular, we compute the 90$\%$ confidence interval of the marginalized posterior of eccentricity to discern at what value of the injected eccentricity the marginalized posteriors transition from clustering mostly near zero to clustering away from zero without considerable support at zero.

Let us describe the confidence interval that we will use in our results. We here work with the minimum volume confidence interval which is obtained by defining a threshold probability ($p_{\ast}$) and integrating those regions where the probability density exceed this limit. This limit is lowered until the total probability reaches the desired amount. This minimum volume confidence interval is given by 
\begin{equation}
\label{eq:min_conf_int}
\int [p(x) \geq p_{\ast}]dx = \alpha \, ,
\end{equation}
where $\alpha$ is the desired credible interval.

Figure \ref{fig:low_e_posts} shows some of the marginalized posteriors obtained for a $(1.4, 1.4)M_{\odot}$ signal with various eccentricities, as well as their 90$\%$ confidence interval, and the value of the injected eccentricity. In the low eccentricity case, the posterior is effectively producing samples of the uniform flat prior in the regime where eccentricity is very small ($e_{\ast} \leq 0.07$), but excludes higher eccentricities. This leads to the lack of structure in the peaks of the posterior at low injected eccentricity. As the injected eccentricity is increased, the posterior still includes zero in the 90$\%$ confidence interval, but there is a more defined peak and the distribution begins to take on a Gaussian shape whose peak is shifted from zero. Finally, as the injected eccentricity is raised high enough, there is little support for zero eccentricity, and the posterior takes on a Gaussian shape with a well defined peak near the injected value (the peak does not lie directly on the injected value due to marginalization along other covariant parameters).

Figure \ref{fig:low_e} shows the maximum of the marginalized posterior (denoted with an``x") of $e_{\ast}$ for a $(1.4, 1.4)M_{\odot}$ system (left) and a $(10, 10)M_{\odot}$ system (right), together with vertical error bars constructed from the 90$\%$ minimal area confidence intervals (shown e.g.,~in Fig.~\ref{fig:low_e_posts} for the NS binary case). In the low and high mass cases, the 90$\%$ confidence begins to exclude zero eccentricity at $e_{\ast, \rm inj} = 0.008$ and at $e_{\ast, \rm inj} = (0.05)$, respectively. These results suggest that we could in principle be detecting even very small eccentricities with advanced LIGO at design sensitivity, if such systems exist in Nature. For lower mass sources, like a neutron star binary, LIGO could measure eccentricities as low as $e_{\ast} \sim 0.01$ and for higher mass systems $e_{\ast} \sim 0.05$. Of course this also is dependent on the SNR of the signal, which here is assumed to be 15. We could also possibly provide a better measurement given an eccentric model that incorporates merger and ringdown effects in the waveform. 

\section{Systematic Parameter Error from Mismodeling}
\label{sec:param_err}

In Section \ref{sec:large_ecc}, we presented the statistical error in measuring the eccentricity and chirp mass of our model given eccentric signals, and tempered these results with the systematic error due to model inaccuracy. In the following sections, we expand on that discussion of systematic error by considering two sources of this error from model inaccuracy: error from neglecting orbital eccentricity altogether in our models when recovering parameters from an eccentric signal, and the error introduced by model inaccuracy in our eccentric model. 

We study systematic error by assuming that the signal is described by an ``exact" model, given by $h_{\rm EX}(\theta^{i})$, and it is fitted by an approximate model, given by $h_{\rm AP}(\theta^{i})$. The model that best fits the signal for a given set of injected parameters is the one that maximizes the likelihood, or by appropriately replacing $\bold{d}$ and $\bold{h}$ in Eq.\eqref{eq:likelihood} minimizes the quantity ($h_{\rm EX}(\theta^{i}_{\rm inj}) - h_{\rm AP}(\theta^{i})|h_{\rm EX}(\theta^{i}_{\rm inj}) - h_{\rm AP}(\theta^{i})$). We refer to the parameters of the model, $\theta^{i}$, which minimize this quantity as the ``best-fit parameters'' $\theta_{\rm bf}^{i}$, and the injected parameters of the ``exact" signal as $\theta_{\rm inj}^{i}$. The systematic parameter error is the difference between the parameters of the signal, and the best-fit parameters of the approximate model, i.e., 
\begin{equation}
\label{eq:param_err}
\delta \theta^{i} = \theta^{i}_{\rm inj} - \theta^{i}_{\rm bf} \, ,
\end{equation}

Clearly then, to study this parameter error and obtain $ \theta^{i}_{\rm bf}$ we require a very precise estimate of the maximum likelihood (which is identical to finding the best fit in our analysis) when injecting an ``exact" signal and recovering with an approximate model. In order to obtain this very accurate maximum likelihood estimation, we use our MCMC algorithm with a signal injected at a very large SNR, so that the likelihood, as a function of the parameters of the approximate model $\theta^{i}$, is very narrowly peaked. We employ many chains so that the coldest chain can effectively explore the very narrowly peaked likelihood. After several thousands of iterations with no new maximum found, we end the search.
\subsection{What Systematic Error is Incurred by Neglecting Eccentricity in Waveform Modeling?}
\label{sec:err_circ}
In this subsection we study the systematic parameter error which results from neglecting the effects of eccentricity in the model. In order to do so, we inject an eccentric signal and recover it using our model with the eccentricity fixed to zero\footnote{In reality the model requires a solution for how the eccentricity changes with orbital separation so we have used a very small eccentricity of $10^{-10}$, which is effectively circular.}, i.e.,$h_{\rm EX}(\theta^{i})$ is the TaylorF2e+ model, while $h_{\rm AP}(\theta^{i})$ is the circular limit of the TaylorF2e+ model. We employ both the method mentioned above to estimate the best-fit parameter values which gauge systematic error, as well as our MCMC algorithm to produce posteriors and extract confidence intervals which gauge statistical error, as we will explain below. We then explore how using the circular model given an eccentric signal leads to biases in the recovery of $\ln \mathcal{M}_c$ exploring both systematics and increase in statistical error.

\begin{figure*}[htp]
\includegraphics[clip=true,angle=0,width=0.475\textwidth]{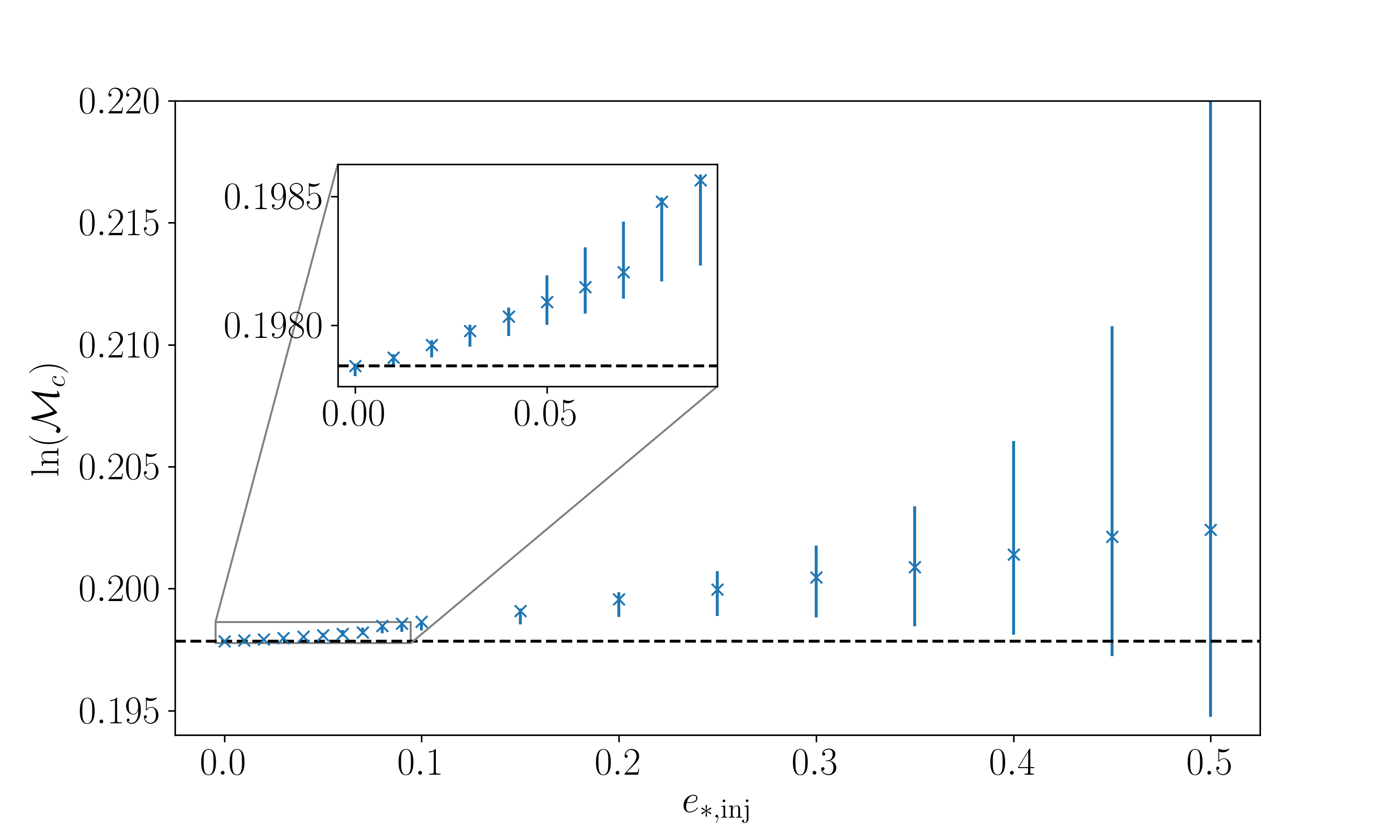}\includegraphics[clip=true,angle=0,width=0.475\textwidth]{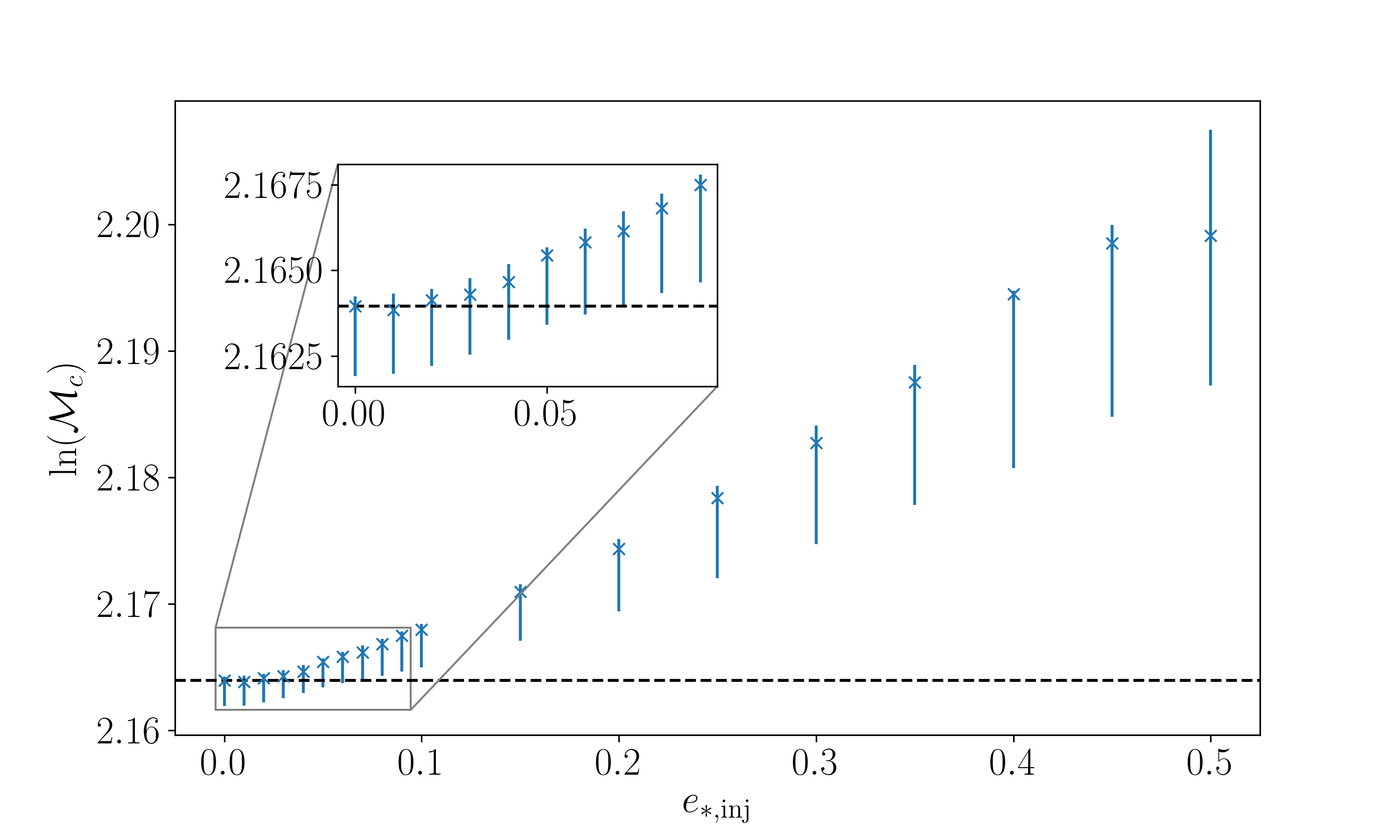}
\caption{\label{fig:circ_bias} Best-fit value of $\ln \mathcal{M}_c$ (``x"), as a function of injected eccentricity, together with a measure of the statistical error in the extraction of this parameter, when using a circular template to recover an eccentric signal. The vertical error bar represents the 90$\%$ confidence interval in the extraction of $\ln \mathcal{M}_c$. The dashed line indicates the injected value of $\ln \mathcal{M}_c$. The left panel corresponds to a $(1.4,1.4)M_{\odot}$ mass injection and the right corresponds to a $(10,10)M_{\odot}$ mass injection, while the SNR of the injected signal is fixed at 15. Observe that the best-fit parameter diverges away from the injected value at injected eccentricities as low as 0.02 (0.08) for the less (more) massive binary.}
\end{figure*}

Figure \ref{fig:circ_bias} shows the best-fit value of $\ln \mathcal{M}_c$ (as an ``x") as a function of injected eccentricity, together with the 90$\%$ central confidence interval in the marginalized posterior on $\ln \mathcal{M}_c$ (vertical error bar). We show results for a $(10,10)M_{\odot}$ and $(1.4,1.4)M_{\odot}$ system. The central confidence interval is computed by integrating the posterior toward its maximum value from each side until the appropriate amount of probability has been collected. Thus, to find the bounds $x_{\rm min}$ and $x_{\rm max}$ of the central confidence interval of size $\alpha$ for a probability distribution $p(x)$ one uses
\begin{equation}
\label{eq:cent_conf_int}
\int^{x_{\rm min}} p(x)dx = \int_{x_{\rm max}} p(x)dx = \frac{\alpha - 1}{2} \, .
\end{equation}

In each case shown in Figure \ref{fig:circ_bias}, the best-fit mass (whose difference from the injected value represents the systematic error) increases and is larger than the injected value, while the confidence region grows, as the injected eccentricity is increased. The bias toward higher mass is, at least in part, due to the fact that an eccentric system inspirals faster than its circular counterpart (depending on the length that is held fixed while taking the circular limit), and more massive systems inspiral faster. For the lower mass system, the 90$\%$ confidence interval in $\ln \mathcal{M}_c$ does not contain the injected value for eccentricities larger 0.02, and for the more massive case for eccentricities larger than 0.08. In each case, the best-fit value of the mass differs from the injected value by the time the eccentricity of the signal reaches $e_{\ast} \sim 0.02$. However, in the low mass case the injected value does begin to fall in the confidence interval for very large eccentricities -- a result of that interval becoming very large. 

This suggests that eccentricity should be considered when carrying out parameter estimation even in cases where the signal could be very nearly circular ($e_{\ast} \sim 0.02$) in order to avoid systematic bias. The value of the injected eccentricity of the signal at which this systematic bias becomes important is unsurprisingly close to the value at which eccentricity becomes measurable, as shown in Sec.~\ref{sec:small_ecc}. Obviously, the degree to which the systematic error exceeds the statistical error will depend on the signal SNR (as systematic error is independent of SNR and statistical error decays as ${\rm{SNR}}^{-1}$), which in these studies was fixed at 15. We also note that these investigations should also be carried out in the future with a model that incorporates merger and ringdown, when the later is developed, as this could affect some of the quantitative details of the conclusions presented above. 

\subsection{What Systematic Error is Incurred by Inaccuracies in the Modeling of Eccentric Binaries?}
\label{sec:err_model}
In this subsection, we illustrate the systematic parameter error in the estimation of the eccentricity induced by mismodeling in our eccentric waveform. In the most ideal situation, we would have access to a catalog of numerical relativity waveforms with eccentricities as high as $e_{\ast} = 0.8$ and durations consistent with our TaylorF2e+ waveform. Unfortunately, these numerical waveforms are costly to compute and very few exist that incorporate eccentricity. However, we do have access to our TaylorT4-like waveform, which is obtained by numerically solving the PN binary dynamics, plugging these into the appropriate PN expression for $h(t)$, and then taking a discrete Fourier transform numerically. Thus, we will treat this numerically-solved, PN model as exact, use it as our true model, and find the TaylorF2e+ that fits it best, i.e.,~$h_{\rm AP}(\theta^{i})$ is the TaylorF2e+ model, while $h_{\rm EX}(\theta^{i})$ is the TaylorT4-like model described above. This provides the best estimate of the systematic error from model inaccuracy of TaylorF2e+ that is currently available to us. As a byproduct of this study, we produce the fitting factor as a function of chirp mass and eccentricity, which is presented in Appendix \ref{app:validation}.

Figure \ref{fig:de} shows $\delta e$ (i.e.,~the systematic error in the estimation of the eccentricity as defined by Eq. \ref{eq:param_err}) as a function of the injected eccentricity and chirp mass. To the right of the blue line, this error exceeds $\sigma_{e_{\ast}}$ for a SNR 15 signal. Generally as the injected eccentricity increases, the systematic error increases which is consistent with the match calculations in \cite{2019CQGra..36r5003M} as well as the matches shown in Figure \ref{fig:match_f2e+}. As the system becomes less massive, the systematic error decreases but our ability to constrain eccentricity increases, so we still require more faithful templates even for these low mass systems. This is especially true as the SNR is increased, because our ability to constrain parameters will also increase, but the systematic error from model inaccuracy will be independent of SNR. We also see that the error is increased for unequal mass systems (as expected from Figure \ref{fig:match_f2e+}).
\begin{figure}[htp]
\includegraphics[clip=true,angle=0,width=0.475\textwidth]{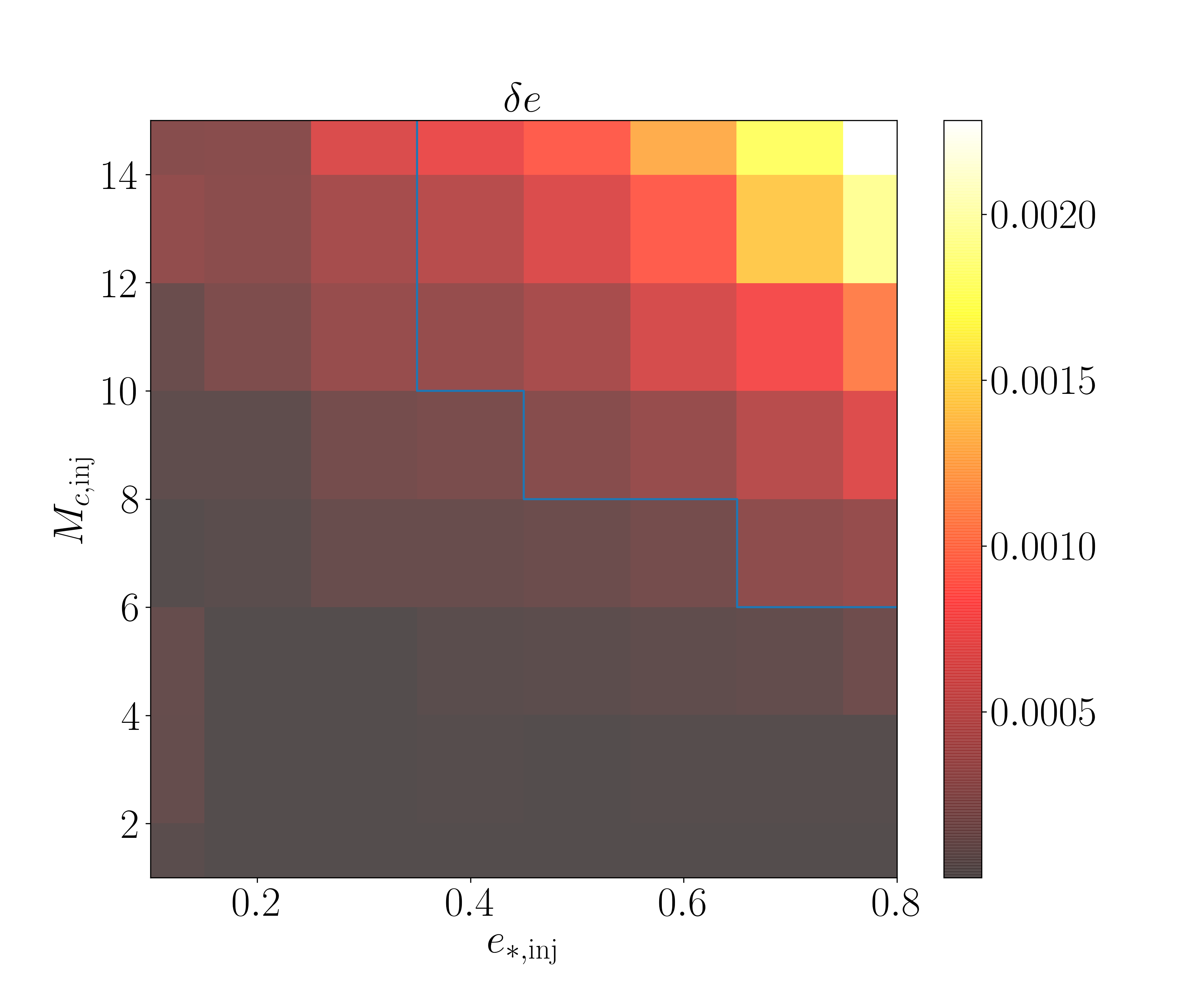}
\includegraphics[clip=true,angle=0,width=0.475\textwidth]{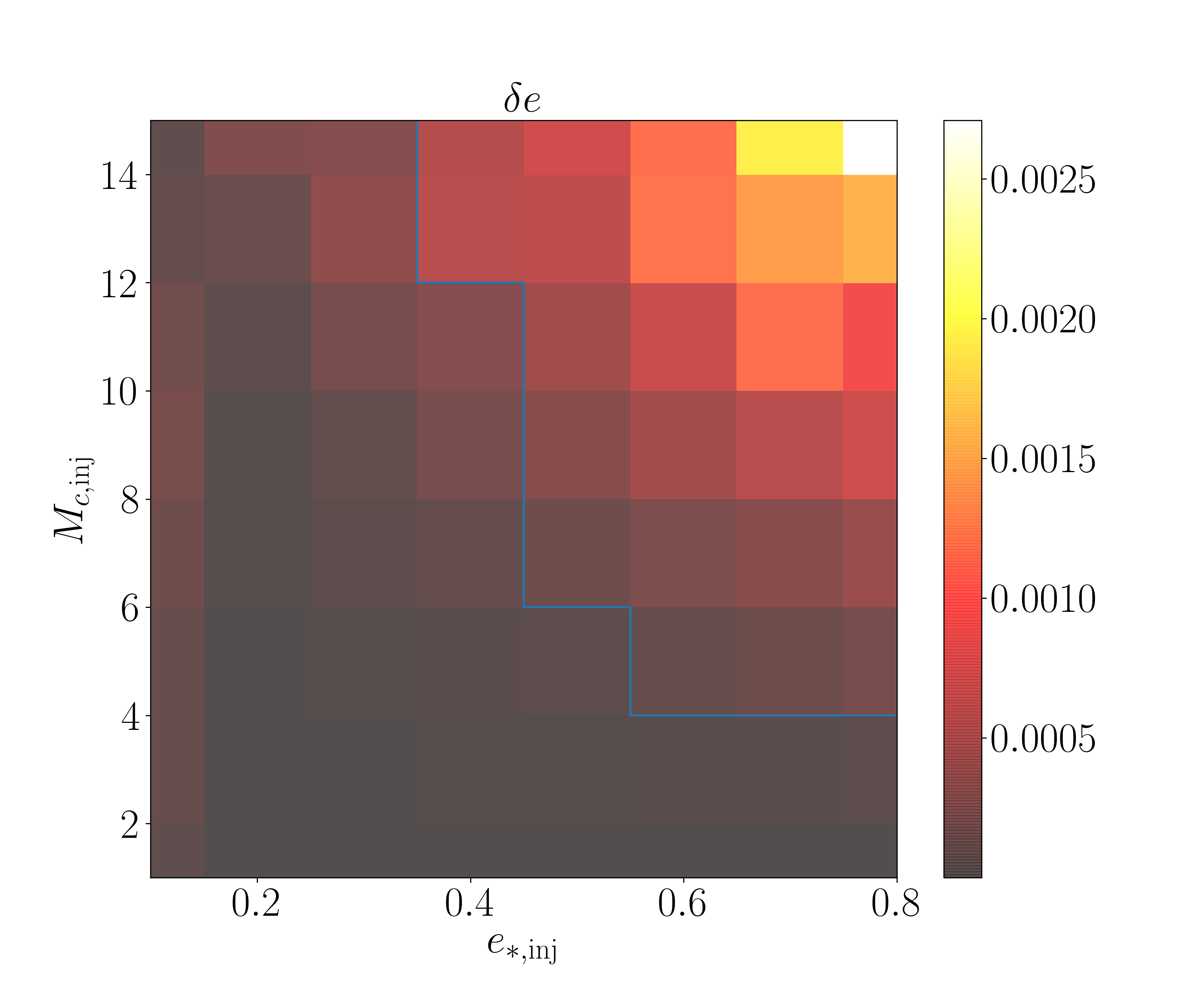}
\caption{\label{fig:de} The systematic error described by Eq. \eqref{eq:param_err}. To the right of the blue line, the systematic error is larger than $\sigma_{e_{\ast}}$ for an SNR 15 signal. Results for $\eta_{\rm inj} = 0.25$ (top) and 0.20 (bottom) are shown.}
\end{figure}

The parameter error seen here suggests that we require better modeling to accurately measure eccentricity, especially for more eccentric signals. In \cite{2019CQGra..36r5003M} we showed results that strongly suggest that the error in match between our TaylorF2e+ model and the fully numerical time-domain model are sourced by PN differences (and not eccentric series expansions). Moreover, this error was seen to become exacerbated by decreasing $\eta$ and increasing eccentricity. This error in the match is what sources the systematic error we see here \cite{2017PhRvD..95j4004C}. This was ultimately illustrated by computing the match with different equally PN valid time-domain waveforms. The results of this portion of our study supports the conclusion of that work: we require higher PN order corrections, or a different and more accurate PN-based model, to faithfully model eccentric sources and extract parameters when the eccentricity becomes large and systems have significantly unequal masses.  

\section{Conclusions \& Future Work}
\label{sec:conc}
We have explored the ability of aLIGO at design sensitivity to measure eccentricity in both the moderate and small eccentricity regime. Generally, as the eccentricity of the signal increases, so does our ability to measure it with $\sigma_{e_{\ast}}$ increasing from $\mathcal{O}(10^{-2})$ for small eccentricities ($e_{\ast} \sim 0.2 $) to about $\mathcal{O}(10^{-4})$ for moderate eccentricities ($e_{\ast} \sim 0.6 $). The accuracy to which the chirp mass can be measured ($\sigma_{\ln \mathcal{M}_c}$) also sees nearly a factor of ten improvement as the eccentricity of the signal increases from 0.1 to 0.4. For nearly-circular signals ($e_{\ast} \leq 0.05$), the eccentricity model parameter begins to become measurable already when $e_{\ast} \sim 0.008$, but this strongly depends on system mass, with lower mass systems having smaller detectable eccentricity thresholds.

Importantly, we discern regions of the ($\mathcal{M}_c, e_{\ast}$) parameter space where the systematic error becomes larger than the statistical error. For low mass systems, the systematic error in our model is sufficiently low even when the eccentricity of the signal is as high as 0.8 (extrapolation of Figure \ref{fig:sigma_heat} suggests that the model could be faithful out to eccentricities as high at 0.9 for a binary neutron star type system). As the signal-emitting system becomes more massive, the systematic error becomes large for eccentricities around 0.5 suggesting more faithful modeling in that regime is required. As the masses of the emitting binary becomes unequal, the systematic error also increases. In \cite{2019CQGra..36r5003M} we presented results that very strongly suggest that the source of this error is ultimately PN disagreement between otherwise equally valid waveforms. Thus, in order to decrease the systematic error in eccentric modeling, a better PN model is required.

As a result of the changes to model parameterization and approximation in the interest of model efficiency, we explored the model's validity in ($\mathcal{M}_c, e_{\ast}$) in terms of the match and fitting factor. We have a model that is very faithful for moderate eccentricities, while still being highly efficient ($\sim$90ms per waveform evaluation). This means our model is realistically useful for data analysis techniques that require the computation of the likelihood many millions of times. 

One large assumption of this work, the neglect of compact object spin, could qualitatively change our results. In the future, inclusion of spin in moderately eccentric waveforms should be considered. In addition, modeling of the merger and ringdown would complete the model and provide a more complete picture of our ability to constrain parameters by making the covariances between all parameters clearer. We have also assumed a single detector in this work and, as a result, we have merged source orientation parameters into a single overall amplitude. Further work should consider a network of detectors and investigate the effect of orbital eccentricity on sky localization.

An interesting extension of this work would be to extend the results of \cite{2019arXiv190807089M}, where they incorporated leading-order eccentricity effects due to a particular scalar-tensor theory of gravity. The effect of eccentricity coupling to this theory leads to improved constraints on the particular theory, however that model was only valid in the small eccentricity regime. Incorporation of those effects into our model which is valid for larger eccentricities could yield an avenue for better constraints on that theory, given moderately eccentric detections. 

Lastly, with the Laser Interferometer Space Detector (LISA) set to launch in the near future and other 3G detectors planned, and the expectation that these detectors will see many eccentric sources, it is highly important to gauge the validity and viability of this eccentric model for such detectors. It is plausible to expect that given these detectors increased sensitivities and diverse frequency bands, we might require even more accurate models, but also an improved ability to measure eccentricity.
  
\section*{Acknowledgments} 

B. M. was supported by the Joan L. Dalton Memorial Fellowship in Astronomy from Montclair State University. B.~M.~and N.~Y.~ also acknowledge support from NSF PHY-1759615 and NASA ROSES grant 80NSSC18K1352. We thank Travis Robson and Neil Cornish for very useful conversations. Computational efforts were performed on the Hyalite High Performance Computing system, which is supported by University Information Technology at Montana State University.

\appendix
\section{Corner Plots}
\label{app:corner}
\begin{figure*}[htp]
\includegraphics[clip=true,angle=0,width=0.45\textwidth]{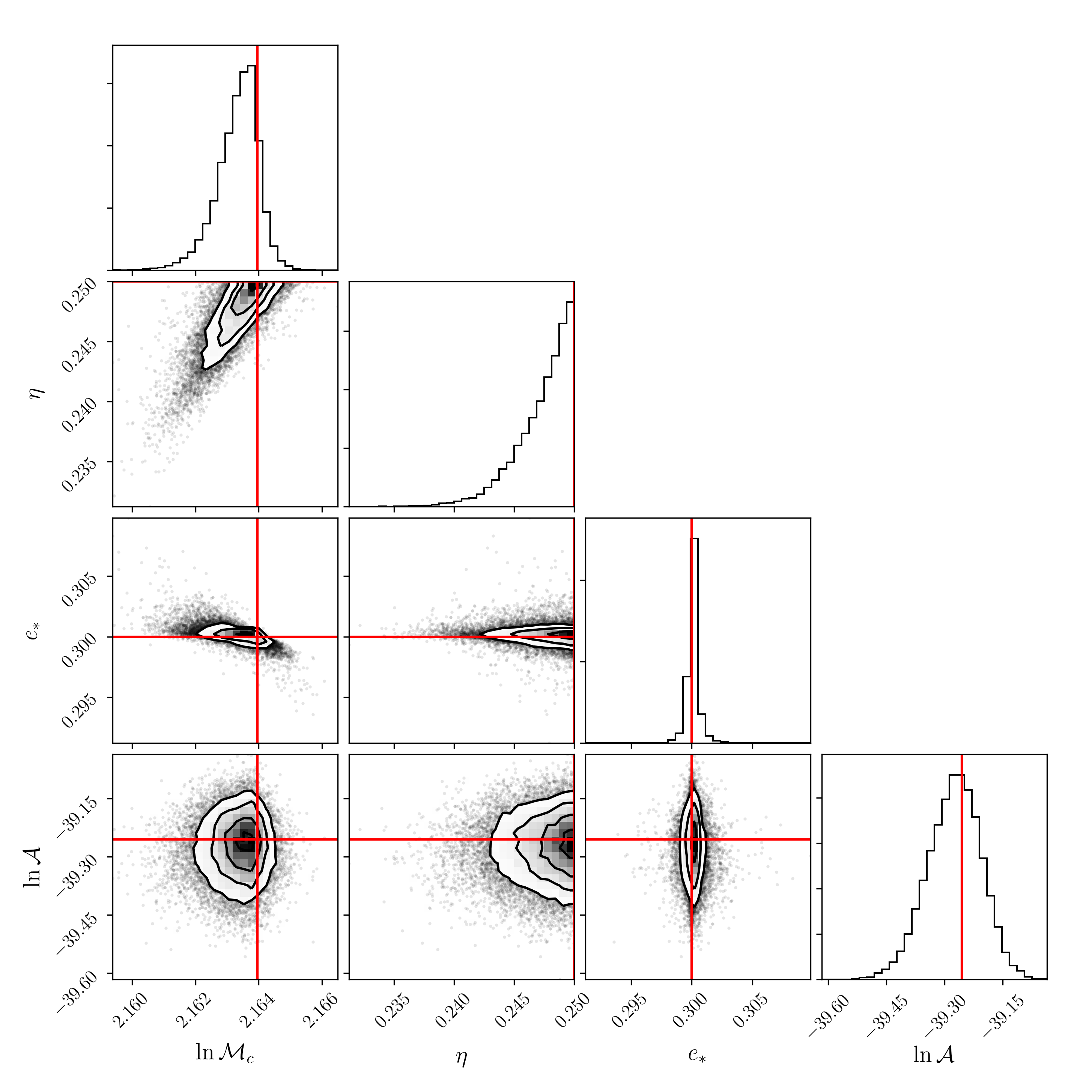}
\includegraphics[clip=true,angle=0,width=0.45\textwidth]{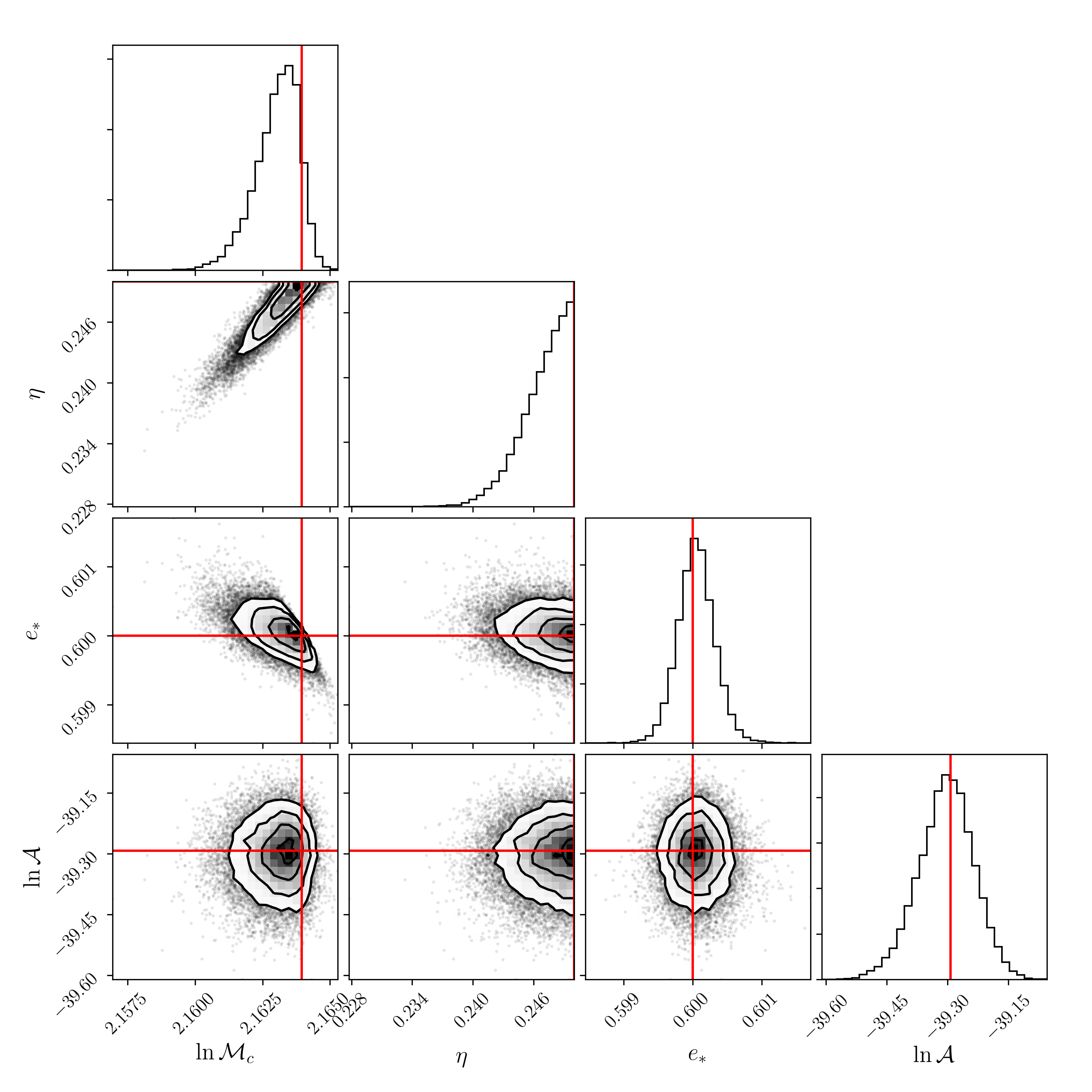}
\includegraphics[clip=true,angle=0,width=0.45\textwidth]{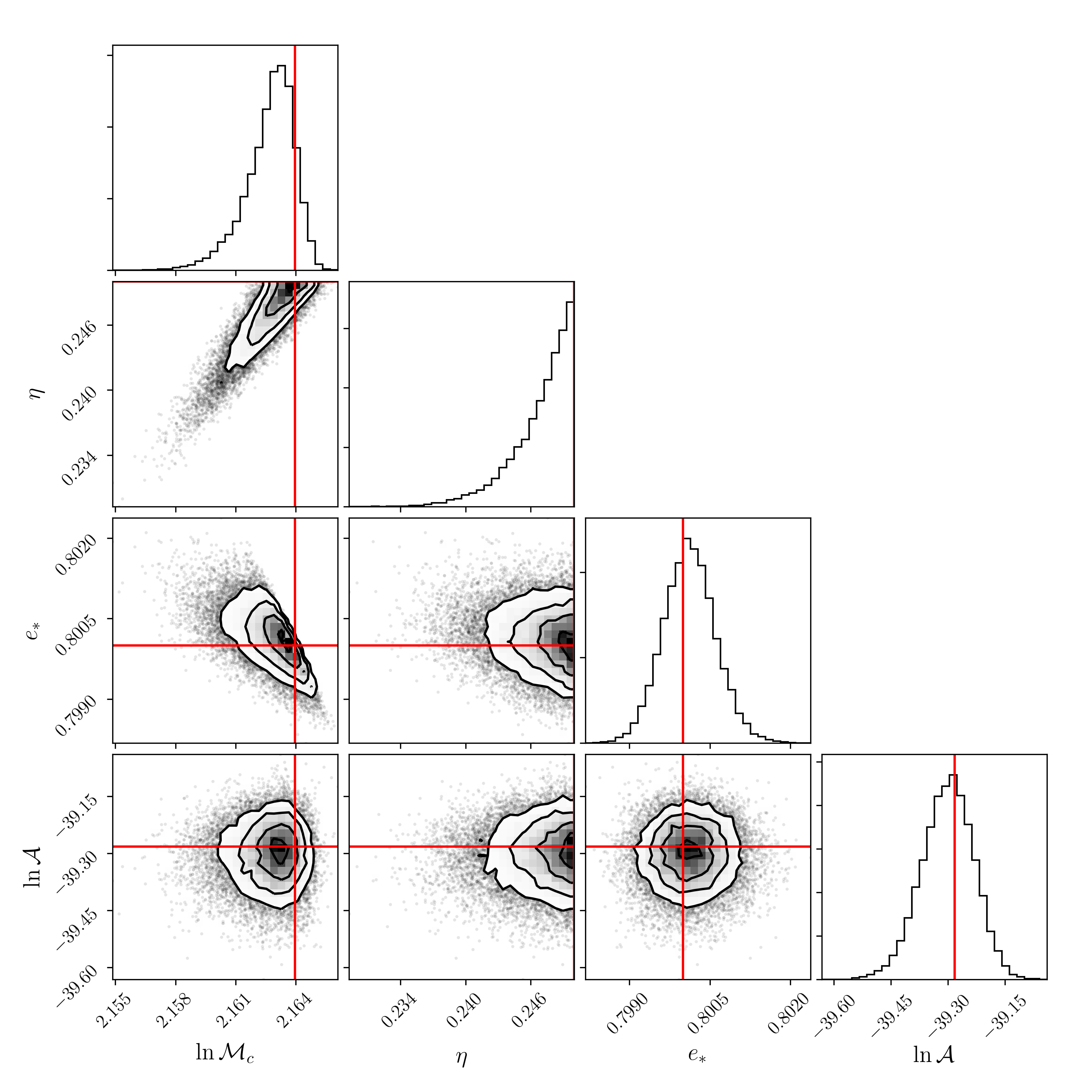}
\includegraphics[clip=true,angle=0,width=0.45\textwidth]{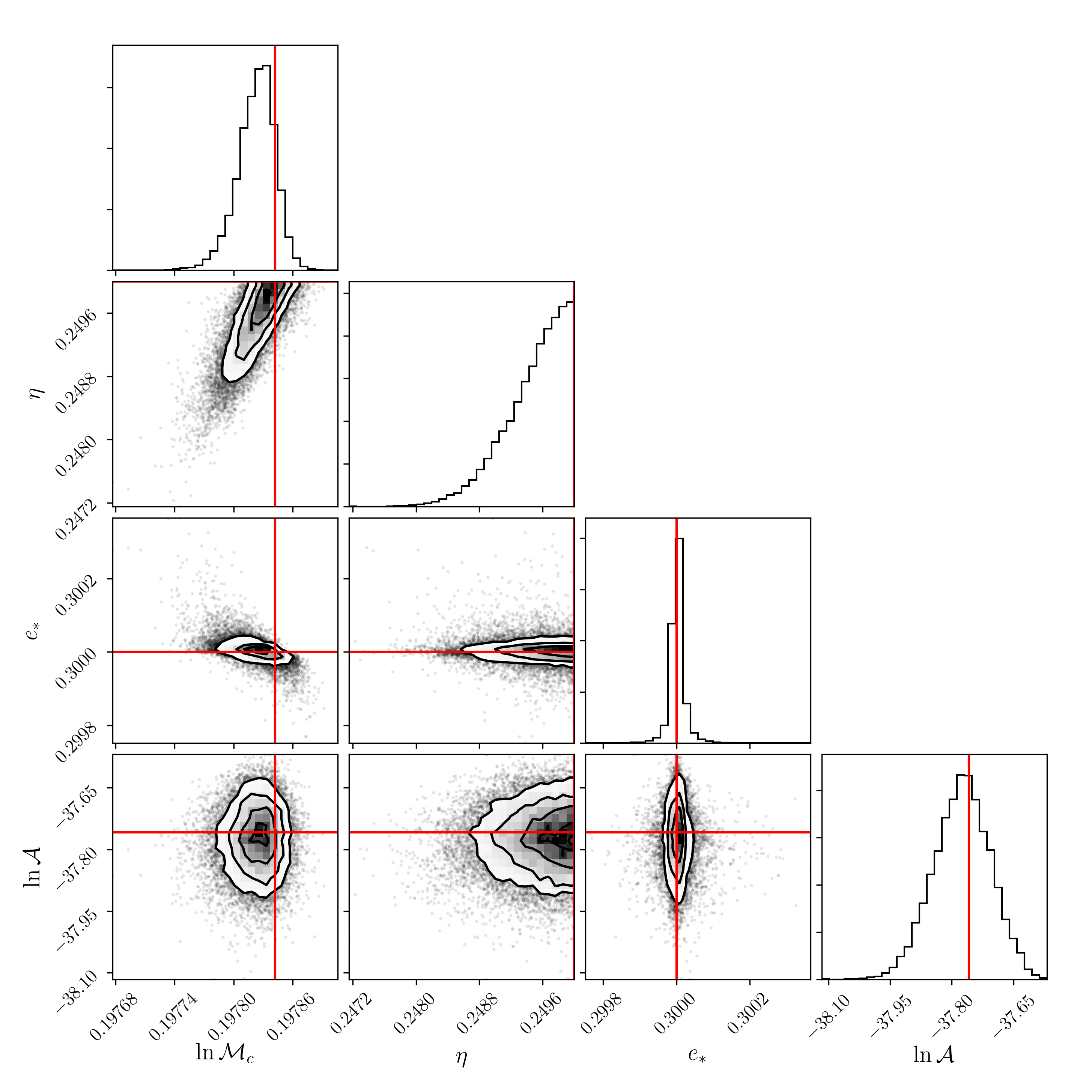}
\caption{\label{fig:corner} Posteriors for a SNR 15 signal with (1.4,1.4)$M_{\odot}$ (bottom right) and (10,10)$M_{\odot}$ (top, bottom left) masses. The red lines indicate injected values. Observe that most of the marginalized posterior appear fairly Gaussian in shape. As the injected eccentricity is increased, a stronger covariance between eccentricity and mass ratio develops. All posteriors are more closely clustered on the injected values in the low mass case.} 
\end{figure*}

In this appendix, we present a few select corner plots to get a complete  picture of the results of other sections, which use the results of 100s of posteriors to paint a general picture of the likelihood surface. On the diagonal of the corner plots is the marginalized posterior, and on the off diagonals are the 2-D joint posteriors, which show the covariances between parameters.

Figure \ref{fig:corner} presents the posteriors for a $(10, 10)M_{\odot}$ mass injection with injected eccentricities of 0.3 (top left), 0.6 (top right), and 0.8 (bottom left). Also shown in the bottom right is a a $(1.4, 1.4)M_{\odot}$ mass injection with injected eccentricity of 0.3. Each has a SNR of 15.  We see that (with all else held fixed), as the eccentricity of the signal is increased, the posterior on $\ln{\mathcal{A}}$ is unchanged. This suggests that measurement of the overall amplitude is mostly independent of eccentricity, though it is worth noting that this study only considers one detector. It is possible that with many detectors, eccentric effects could improve sky localization. As expected from the results of Sec.~\ref{sec:large_ecc}, the marginalized posteriors on eccentricity become more constraining as the eccentricity of the signal is increased, as well as when the mass is decreased. Interestingly, as the eccentricity of the signal increases, there is an increased covariance with chirp mass and eccentricity, as well as covariance with reduced mass ratio and eccentricity.

\section{Matches and Fitting Factors}
\label{app:validation}

In this appendix, we provide a few more measures of faithfulness for our two Fourier domain eccentric waveforms: TaylorF2e+ and TaylorF2e-, which are discussed in Sec.~\ref{sec:model}. We present the match and the fitting factor. The match is the waveform overlap maximized over extrinsic parameters. The fitting factor (FF) is the overlap maximized over all parameters (this is computed using the method to estimate maximum likelihood presented in Section \ref{sec:param_err}).

\begin{figure*}[htp]
\includegraphics[clip=true,angle=0,width=0.475\textwidth]{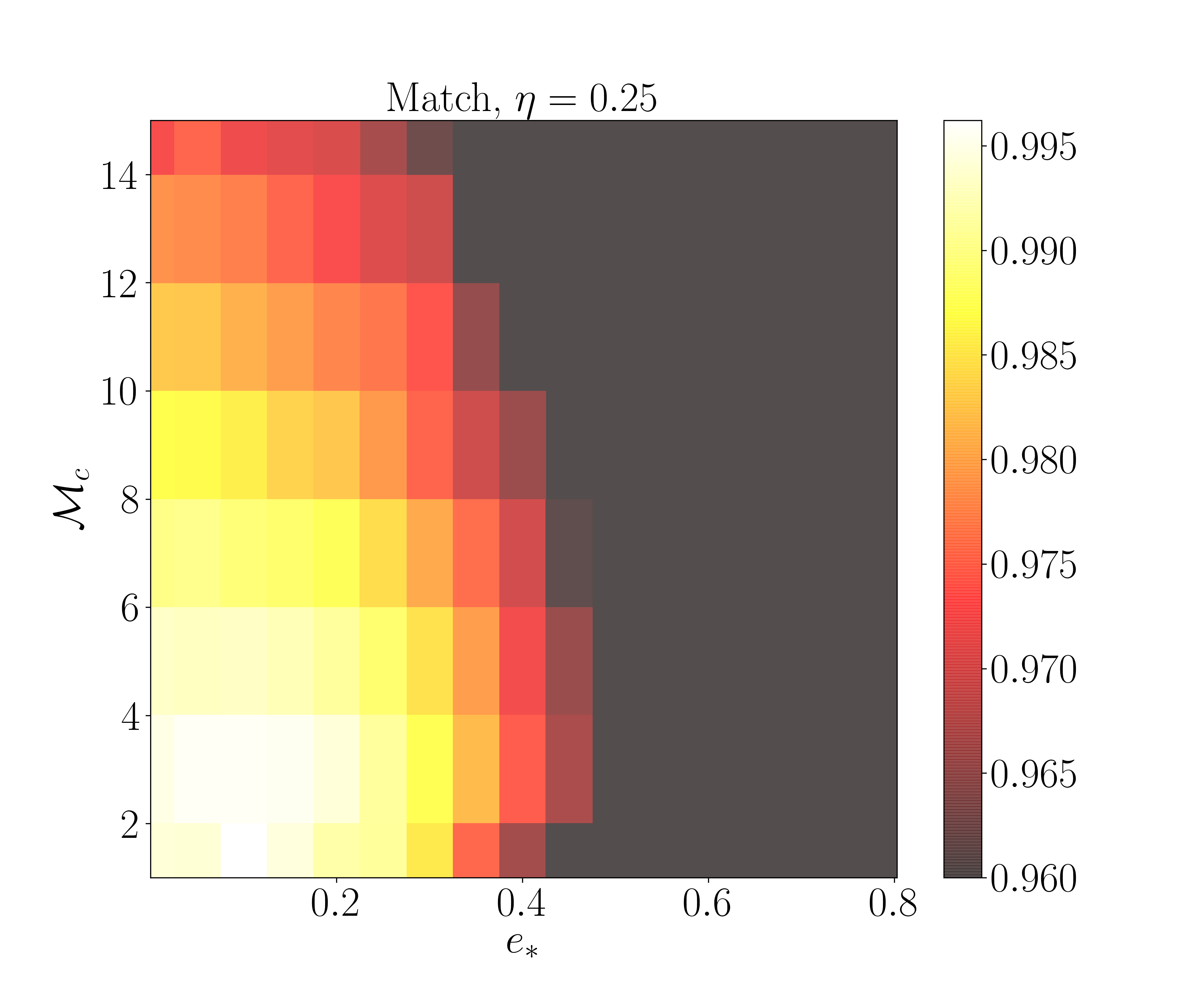}
\includegraphics[clip=true,angle=0,width=0.475\textwidth]{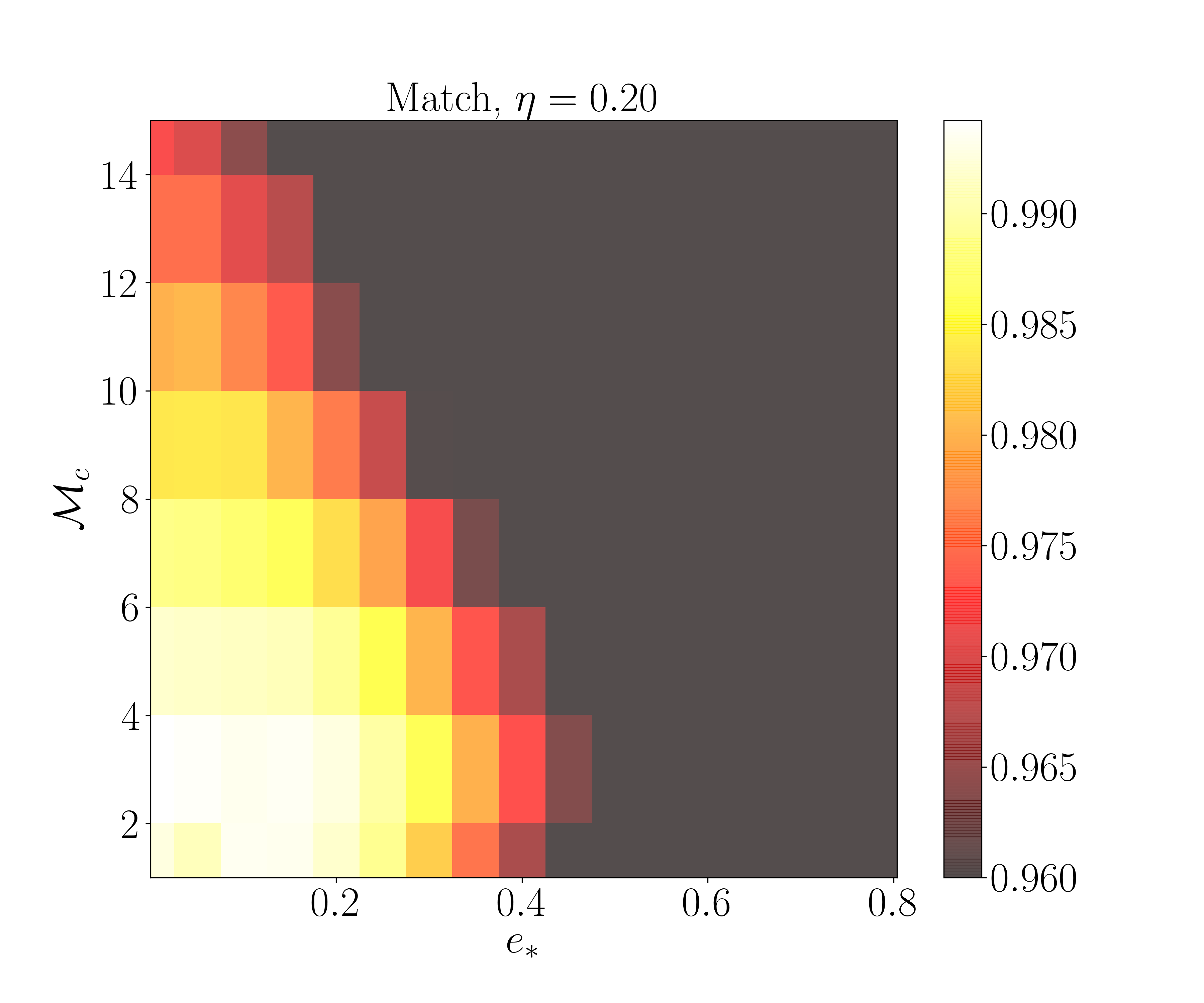}
\caption{\label{fig:match_f2e-} The match between the TaylorF2- model and a fully numerical time-domain model as a function of the chirp mass and $e_{\ast}$ for $\eta = 0.25$ (left) and $\eta = 0.2$ (right). Values below $96\%$ are excluded. Observe that in the low mass cases ($\mathcal{M}_c < 6$), this version of the model performs significantly worse than its TaylorF2e+ counterpart (which keeps more terms in various expansions in eccentricity) as shown in Figure \ref{fig:match_f2e+}.}
\end{figure*}

Figure \ref{fig:match_f2e-} shows the match between TaylorF2e- (our Fourier domain model which keeps less terms in the eccentric expansions appearing in the phases) and a fully numerically solved TaylorT4-like PN model. We see that this model has a match similar to the analogous result for the TaylorF2e+ model (which keeps more terms in eccentric expansions) as shown in Fig.~\ref{fig:match_f2e+}. However, as the mass becomes small, TaylorF2e- is much less faithful, valid only for about half the eccentricities of the TaylorF2e+ model. This suggests that the TaylorF2e+ model should be used if trying to detect or measure parameters for low mass sources. 

\begin{figure*}[htp]
\includegraphics[clip=true,angle=0,width=0.475\textwidth]{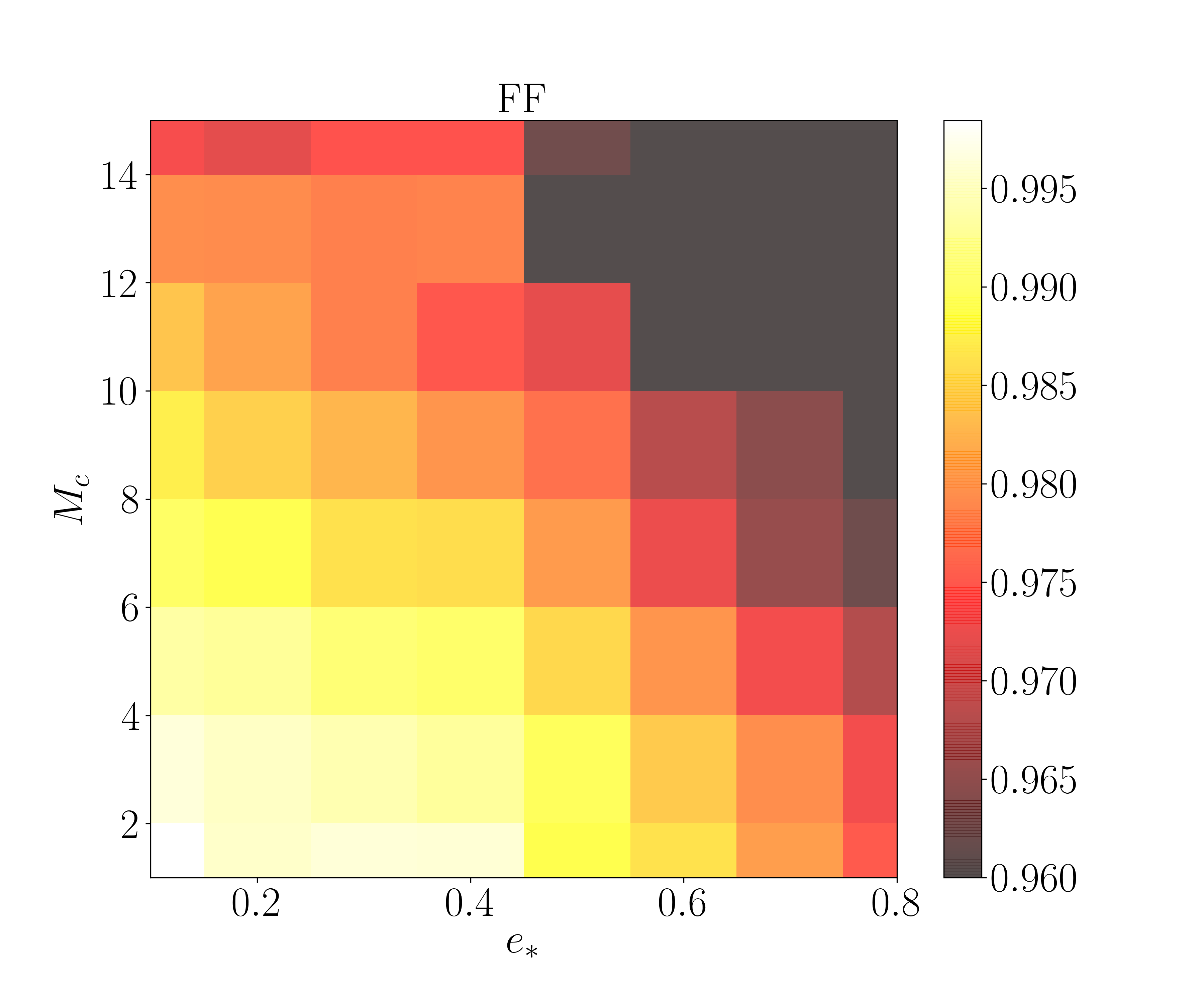}
\includegraphics[clip=true,angle=0,width=0.475\textwidth]{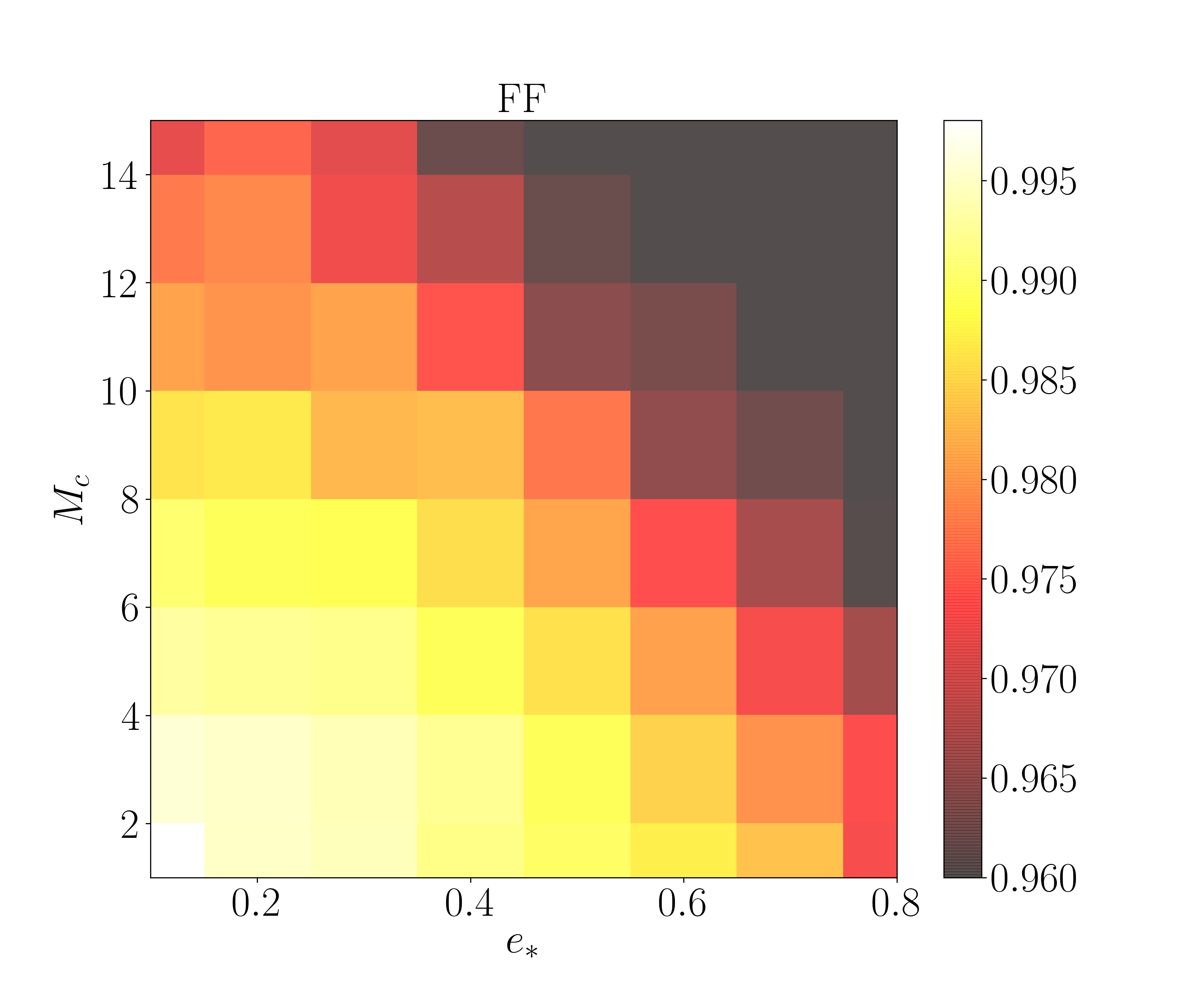}
\caption{\label{fig:FF_f2e+} The fitting factor (FF) between the TaylorF2+ model and a fully numerical time-domain model as a function of the chirp mass and $e_{\ast}$ for $\eta = 0.25$ (left) and $\eta = 0.2$ (right). Values below $96\%$ are excluded. Observe that there is little difference in fitting factor between the equal and unequal mass cases. The fitting factor is greater than $97\%$ for much of the explored parameter space.}
\end{figure*}

Figure \ref{fig:FF_f2e+} shows the fitting factor between the TaylorF2e+ model and our fully numerically-solved, time-domain PN model. We see that the fitting factor is particularly high even for very large eccentricities, provided that the system is low mass. Interestingly, there is little difference between the equal and unequal mass case (unlike in the match). The low fitting factors for high mass and eccentricity again points to the need for more accurate templates in this regime.

\bibliography{master}
\end{document}